\documentclass{article}

\usepackage{arxiv}

\usepackage{amsmath,amssymb,amsfonts}
\usepackage{graphicx} 

\usepackage{textcomp}
\usepackage{color}
\usepackage{bm}
\usepackage{subcaption,mwe}
\usepackage{float}

\usepackage{booktabs}

\usepackage[utf8]{inputenc} 
\usepackage[T1]{fontenc}    
\usepackage{hyperref}       
\usepackage{url}            
\usepackage{booktabs}       
\usepackage{amsfonts}       
\usepackage{nicefrac}       
\usepackage{microtype}      
\usepackage{lipsum}		
\usepackage{graphicx}
\usepackage{natbib}
\usepackage{doi}

\title{Low count of optically pumped magnetometers furnishes a reliable real-time access to sensorimotor rhythm}

\author{{\includegraphics[scale=0.06]{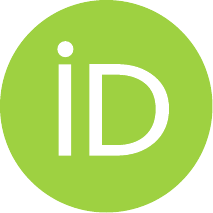}\hspace{1mm}Fedosov Nikita} \\
	Centre for Bioelectric Interfaces \\ Higher School of Economics, Moscow, Russia \\
	\texttt{np\_fedosov@list.ru} \\
\And
\href{https://orcid.org/0000-0000-0000-0000} {\includegraphics[scale=0.06]{orcid.pdf}\hspace{1mm}Daria Medvedeva}\\
	Centre for Bioelectric Interfaces \\ Higher School of Economics, Moscow, Russia\\
LIFT, Life Improvement by Future Technologies Institute\\ Moscow, Russia\\
	\texttt{dariamedvedeva777@gmail.com} \\
	\And
	\href{https://orcid.org/0000-0000-0000-0000}{\includegraphics[scale=0.06]{orcid.pdf}\hspace{1mm}Oleg Shevtsov} \\
	Centre for Bioelectric Interfaces \\ Higher School of Economics, Moscow, Russia\\
LIFT, Life Improvement by Future Technologies Institute\\ Moscow, Russia\\
	\texttt{olegshevts@gmail.com} \\
	\And
	\href{https://orcid.org/0000-0000-0000-0000}{\includegraphics[scale=0.06]{orcid.pdf}\hspace{1mm}Alexei Ossadtchi}\\
	Centre for Bioelectric Interfaces \\ Higher School of Economics, Moscow, Russia\\
LIFT, Life Improvement by Future Technologies Institute\\ Moscow, Russia\\
	\texttt{ossadtchi@gmail.com} \\
}

\renewcommand{\shorttitle}{\textit{arXiv} Template}

\definecolor{forest_green}{RGB}{34, 139, 34}
\definecolor{sky}{RGB}{108, 150, 255}

\hypersetup{
pdftitle={The Optically-Pumped Magnetometers Sensorimotor Rhythm Spectroscopy and BCI application},
pdfsubject={q-bio.NC, q-bio.QM},
pdfauthor={David S.~Hippocampus, Elias D.~Striatum},
pdfkeywords={First keyword, Second keyword, More},
}

\begin{document}
\maketitle

\begin{abstract}
{This study presents an analysis of sensorimotor rhythms using an advanced, optically-pumped magnetoencephalography (OPM-MEG) system—a novel and rapidly developing technology. We conducted real-movement and motor imagery experiments with nine participants across two distinct magnetically-shielded environments: one featuring an analog active suppression system and the other a digital implementation. Our findings demonstrate that, under optimal recording conditions, OPM sensors provide highly informative signals, suitable for use in practical motor imagery brain-computer interface (BCI) applications. We further examine the feasibility of a portable, low-sensor-count OPM-based BCI under varied experimental setups, highlighting its potential for real-time control of external devices via user intentions.}
\end{abstract}


\section{Introduction}

Magnetoencephalography (MEG) is a contemporary and remarkably efficient technique for non-invasive imaging of brain electrical activity offering outstanding opportunities for imaging brain function. The brain magnetic field measuring requires an extremely sensitive and technological tools due to the ultralow magnitude of the magnetic manifestation of neuronal population. This brings on the scene the quantum technologies, including the SQUID-technology most of the modern MEG-devices are based on. These systems have revolutionaized brain imaging and neurological decease diagnostics especially epilepsy. However, being bulky, expensive to purchase and maintain, requiring continuous liquid helium supply and using  non-reconfigurable sensor arrays, the SQUID-based MEG systems did not manage to explore all the treasures conserved in weak magnetic fields generated by neuronal sources. The new hope to non-invasively achieve millimeter and millisecond scale neuroimaging is put now in the modern wearable MEG sensor technology using optically-pumped magnetometers (OPM) \cite{alexandrov2003recent, osborne2018fully,limes2020portable} and other types of highly sensitive wearable magnetic field sensors \cite{koshev2021evolution}. These systems allow for building reconfigurable sensor arrays to fit individual head shapes and the task at hands \cite{petrenko2021towards}.

The OPM-sensors were used in different scenarios and already proved their applicability and reliability in a range of MEG studies, while providing new opportunities - more comfortable and ecological recording conditions including the registration of brain's electromagnetic activity in mobile subjects and children. OPM-sensors with a sensitive element located only 6 mm away from the scalp furnish higher signal-to-noise ratio (SNR) as compared to the SQUID-based systems whose sensors are at least 3 cm away from the scalp due to the thick walls of the Dewar vessel enclosing the liquid helium \textcolor{red}{ \cite{Ivanaynen_}}. The increased SNR opens up a great potential to further boost the spatial resolution of MEG brain imaging modality, which in turn is fundamentally superior to the EEG due to the transparency of the head tissues to the magnetic field and the associated ease and accuracy of forward modelling \cite{singh2014magnetoencephalography}. 

The practical applications of the wearable OPM-MEG span medical fields such as mapping epileptic activity \cite{pedersen2022wearable}, fetal brain activity registration \cite{frohlich2023not}, measuring peripheral nervous system activity \cite{bu2022peripheral}, and contactless measurements of retinal activity \cite{westner2021contactless}, as well as purely neuroscientific endeavors. OPMs also enable new possibilities for research into deep brain structures, such as the hippocampus and cerebellum \cite{tierney2021mouth}, and facilitate the study of the motor system when the subject is in motion \cite{boto2018moving, mellor2023real}. The high accuracy and increased spatial resolution of OPMs, compared with EEG and SQUID-based MEG, enable the classification of more complex signals and significantly advance the development of non-invasive Brain-Computer Interface (BCI). 

These BCI systems translate measurements of cortical signals into control commands \cite{wittevrongel2021practical, shishkin2022active, corsi2024measuring}, and are based on the different types of the signal, from the steady-state visual evoked potentials to the sensorimotor rhythm (SMR)-based, or motor imagery (MI) BCI. {In \cite{zerfowski2021real}, researchers investigated the applicability of OPM sensors for BCI interfacing based on evoked fields, including N200, P300, and SSVEP. They demonstrated robust detection of these fields and the potential to classify commands for external devices, such as a mind-speller. The OPM-SSVEP-based paradigm was further examined in \cite{ji2024user}, which showed a high information transfer rate. For BCIs based on induced activity, there was an attempt to develop an SMR-based BCI in \cite{paek2020towards}, though it focused primarily on actual movement. Another study \cite{zerfowski2021real} used signals from the occipital cortex to classify eye closure and explored applying this pipeline to classify motor imagery from SMR rhythms.}

At the same time, the MI-based BCI is the only true BCI in a sense that it decodes the endogeneously generated user's brain state and do not require any external stimulation. Noteworthy is the utility of such systems in the post-stroke rehabilitation \cite{biryukova2021neurorehabilitation, frolov2017post, paek2020towards}. These systems benefit from the increased SNR and the spatial resolution in measuring brain's rhythmic activity which would allow for the design of more informative systems and more detailed targeting of pathology with advanced clustering and decomposition methods  \cite{frolov2020using}.

Despite the various types of brain activity recorded with OPM sensors to date including attempts to do a BCI, there is limited literature on OPM-based BCI systems operating in a single-trial mode, and, there is a lack of systematic analysis of properties of SMR-rhythm measured by OPM-sensors and its comparison into the MI and real movement (MR) conditions as well as a weak evidence of its usage for the MI-BCI \cite{paek2020towards, fedosov2021motor}.

In this study we carefully and rigorously research the OPM-sensor capabilities and signal features in application the SMR-rhythm recordings and build offline as well as online OPM-BCI setup. Our research aimed to demonstrate that OPM system can measure voluntary modulations of oscillatory activity over the motor cortex and classify the measured data in real-time. These modulations, known as event-related desynchronization/synchronization (ERD/ERS) of the sensorimotor rhythm (SMR), have been used to control external devices in the past. In our setup, the ERD patter was induced by the visual cues promoting them to do a relaxation task and MI or MR tasks. We show the strong ERD patterns for all of the subjects for the MR paradigm as well as for most of them into the MI paradigm. Then we do an offline-classification task with a classic CSP plus LDA pipeline, show, the great potential for the decoding of the imaginary intentions and then demonstrate its applicability into the real-time BCI.

The measuring magnetic brain activity, especially for the single-trial mode classification tasks, requires a very special condition on the environmental noise. The typical range of the brain signals (near the scalp) varies from tens of fT up two 1-2 pT, while the typical values of the slowly-changing Earth magnetic field is around 40 $\mu T$ and urban noise also can reach up to several $\mu T$ depending on the location, which makes it impossible to record brain signal in open environment. To reduce the dynamical range of the magnetic signals the magnetically-shielded rooms (MSR) are used which are the crucial component in zero-field MEG setup and shields most of the detrimental interfering signals. At the same time typical MSR works well for the DC components of the magnetic field and high-frequency oscillations (10 Hz and higher), but near the DC frequency range its performance dramatically drops.

For the good MSR-room it suppresses external DC field about to 50 nT while ultralow frequency components can fluctuate in a range up to $\pm$ 20 nT. The zero-field OPM-sensors which were used in our experiments are highly susceptible to such fluctations as they have vary narrow transfer function and can only measure magnetic field up to several nT. The fluctuations near the measurement limits can also induce some nonlinearities. To solve the problem additional active supression systems based on the wiring coils are usually used. Such systems can vary into the design of the coils and the control system (e.g. digital or analog) and can potentially influence the results of experiment.

We ran our SMR-rhythm experiments in two MSR-rooms, one in the MEG-Center, Moscow and second in a HSE University, Moscow, each with its own type of the interference suppression system. The first one has its own built into the walls analog suppression system with a reference sensor located on a meridian outside the room. The second room does not have the wall-mounted coils and uses a digital adaptive suppression system (DASS) developed by us to ensure a reliable operation of the OPM sensors.  We use this opportunity to present the results of motor-imagery BCI experiments and illustrate how both types of the suppression systems effect on the BCI's performance.

The structure of the results report is as follows: first we describe a DASS which makes the conditions for the reliable BCI-experiments, show the stability of the OPM-sensor measurements. Then, we provide the results for the SMR record in both movement imagery and real movement paradigms, analyse the spatio-temporal properties of the SMR and its modulation. Then, we present results for the robust offline classification of the MI moments versus the Resting states, and show the influence of sensors configuration on it. Finally, we demonstrate an example of successful phantom hand control with the online MI-BCI. Our research demonstrates that an OPM sensor array, in combination with a custom active external field suppression system and data analysis, provides sufficient signal quality and stability for the real-time single-trial decoding required in the context of a BCI application.

\section{Methods}
\subsection{OPM-sensors and magnetically-shielded room}

\begin{figure}[H]
\centering
\begin{subfigure}[t]{.45\textwidth}
    \centering
    \includegraphics[width=\textwidth]{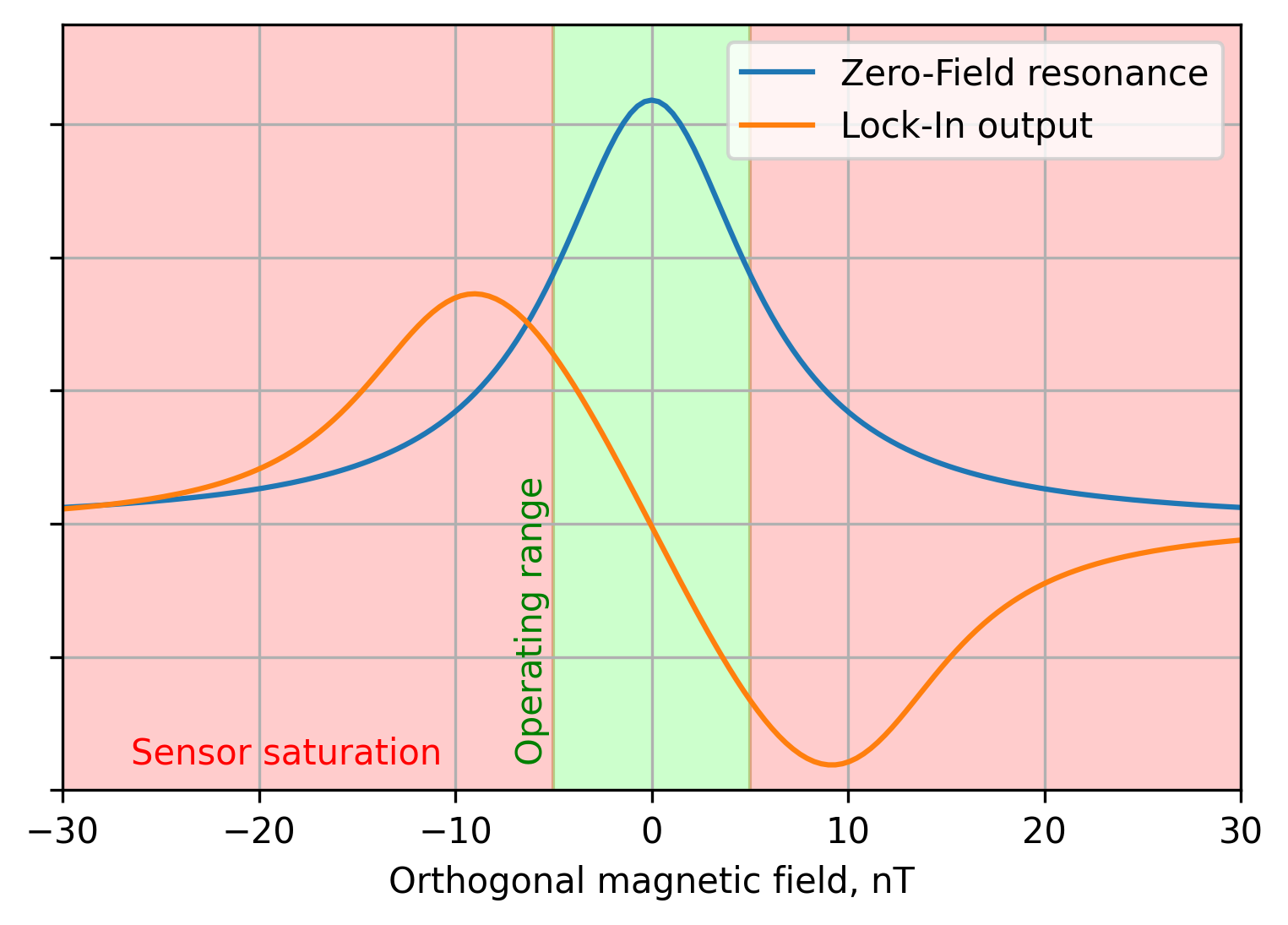}
    \caption{}
    \label{fig:curves} 
\end{subfigure}
\begin{subfigure}[t]{.45\textwidth}
   \centering
   \includegraphics[width=\textwidth]{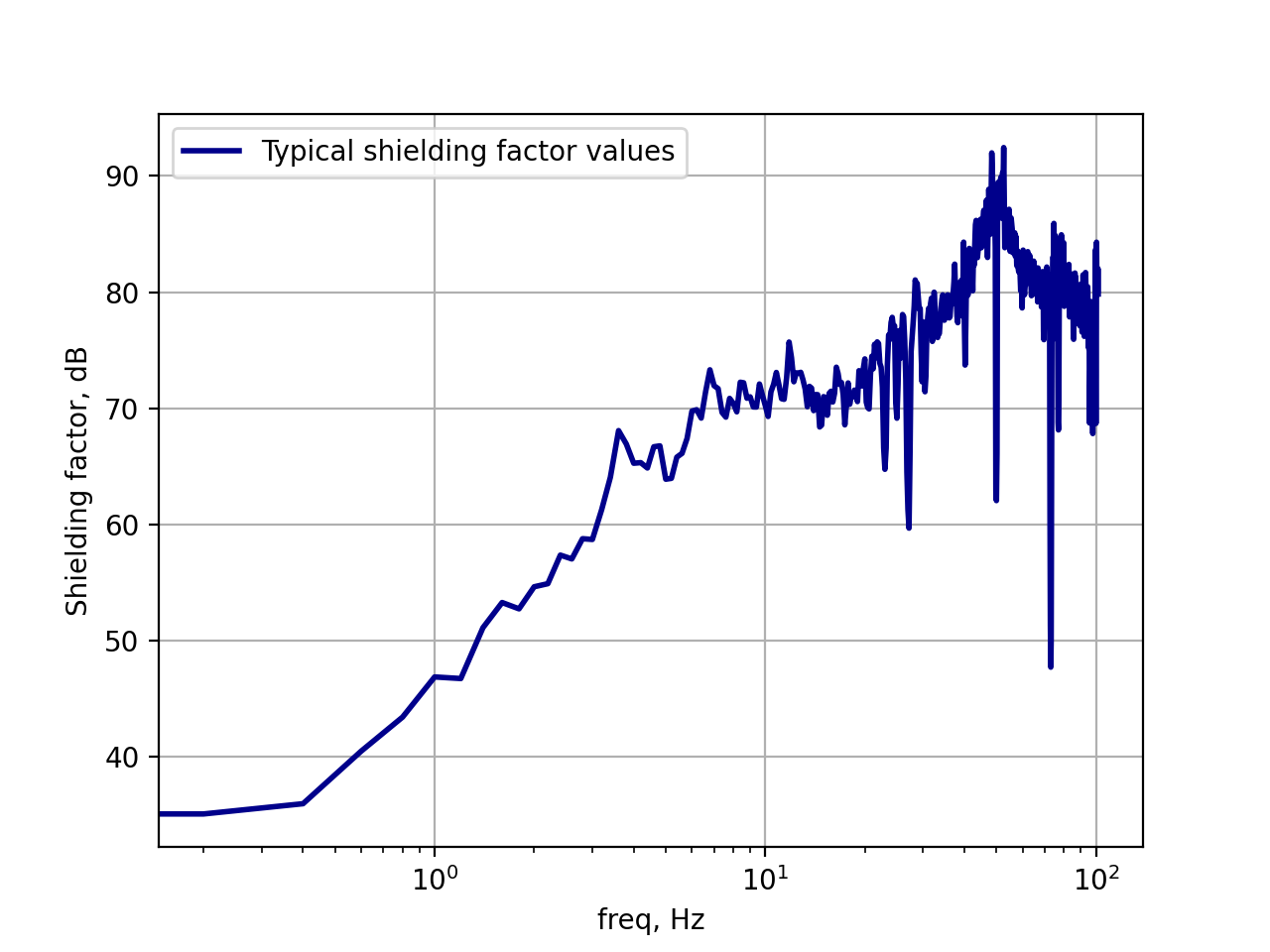}
   \caption{}
   \label{fig:msr}
\end{subfigure}
\caption{a) - Operational curves of the QuSpin OPM sensors. FWHM - full width at half maximum \cite{osborne2018fully}; b) - The shielding factor of the MSR.}
\label{fig:opm_msr_curves}
\end{figure}

{In our research, we used commercially available zero-field sensors, specifically the QuSpin Gen-2 and Gen-3 models (https://quspin.com/products-qzfm-gen2-arxiv/), utilizing up to 7 sensors in total. Each sensor is capable of measuring magnetic fields along two orthogonal axes, with suppression of static magnetic field up to 200 nT, a dynamic range of $\pm$5 nT, and a sensitivity of $<15fT/\sqrt{Hz}$  within the 3–100 Hz frequency band. These sensors exhibit exceptional sensitivity to weak magnetic fields, though their linear response is restricted to a narrow portion of their operational range \ref{fig:opm_msr_curves}. Moreover, the magnetic shielding factor of the MSR shows a pronounced reduction in the ultra-low-frequency range, necessitating supplementary techniques to eliminate residual magnetic fields.}

{The initial environment for conducting our OPM-experiments was the Vacuumschmelze AK-3B MSR, located in the Moscow MEG Center (www.megmoscow.ru/en/), housed within the Moscow State University of Psychology and Education (MSUPE). We will henceforth refer to this experimental environment as Room 1. To address the reduction in shielding efficiency at ultra-low frequencies ($f<$ 0.1 Hz) (see Figure  \ref{fig:curves}), this room is equipped with active shielding coils embedded within the walls. These coils are powered by a high-fidelity, low-noise analog amplifier and a low-pass filter, which together serve to scale and invert the signal from a reference sensor positioned 10–20 meters outside the MSR.}

\begin{figure}[h]
	\begin{center}
		\begin{tabular}{c}
			\includegraphics[width=0.95\linewidth]{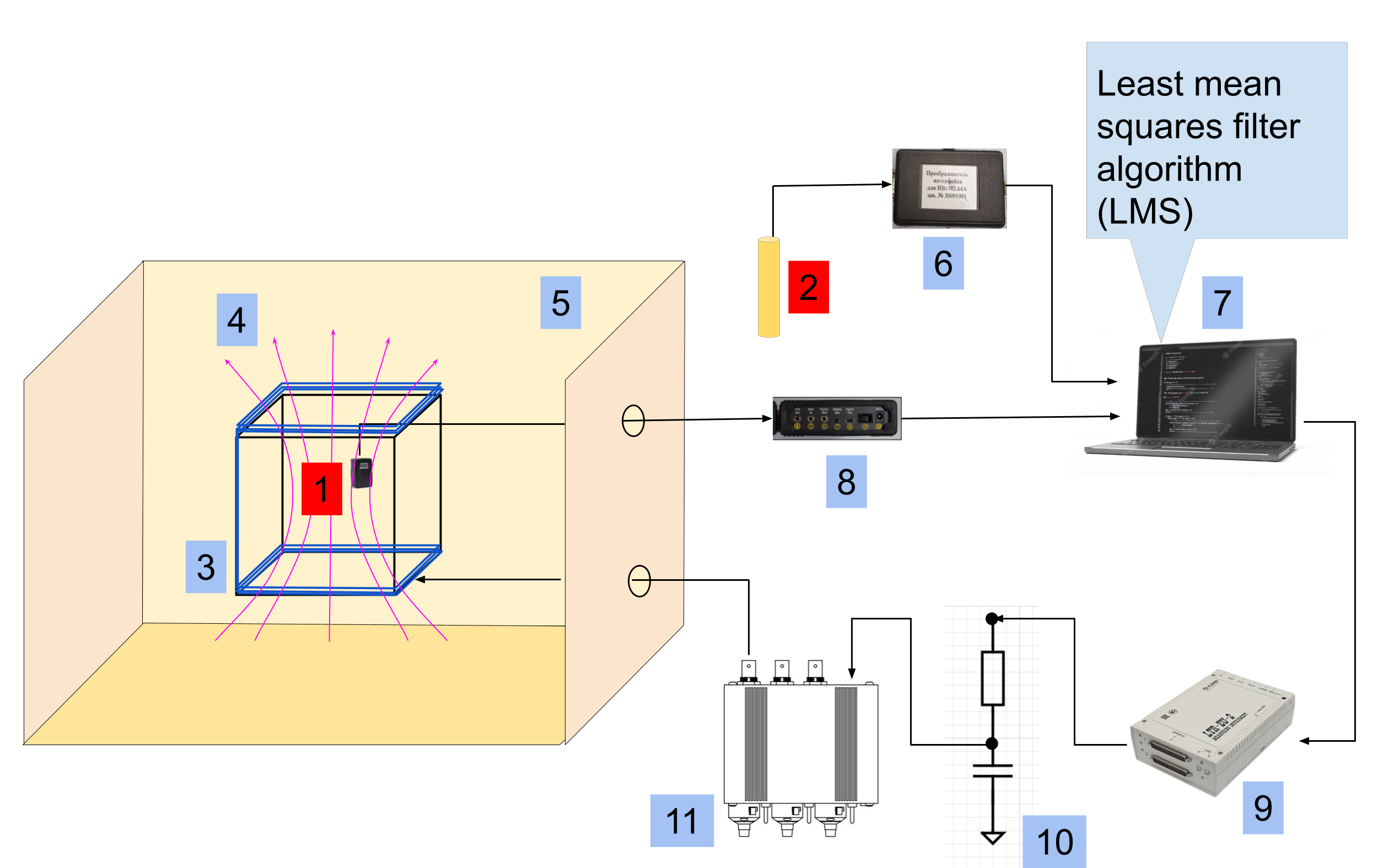} \\
		\end{tabular}
	\end{center}
	\caption{Operation diagram of the adaptive system for suppression of Earth's fluctuating magnetic field: 1 -- the error-measurement OPM sensor; 2 -- the outer sensor (low fidelity inductive sensor with resolution of 3 nT); 3 -- carbon framework with compensation coils arranged in Helmholtz's configuration; 4 -- magnetic flux generated by compensation coils (shown only for the horizontal pair of coils); 5 -- magnetically shielded room; 6 - data registration block for the low fidelity inductive sensor; 7 -- PC; 8 -- OPM data registration block; 9 digital-to analog converter; 10 -- control signal low-pass filter; 11 -- compensation coil driver.}
	\label{fig:asas}
\end{figure}

\begin{figure}
	\begin{center}
		\begin{tabular}{c}
			\includegraphics[width=0.7\linewidth]{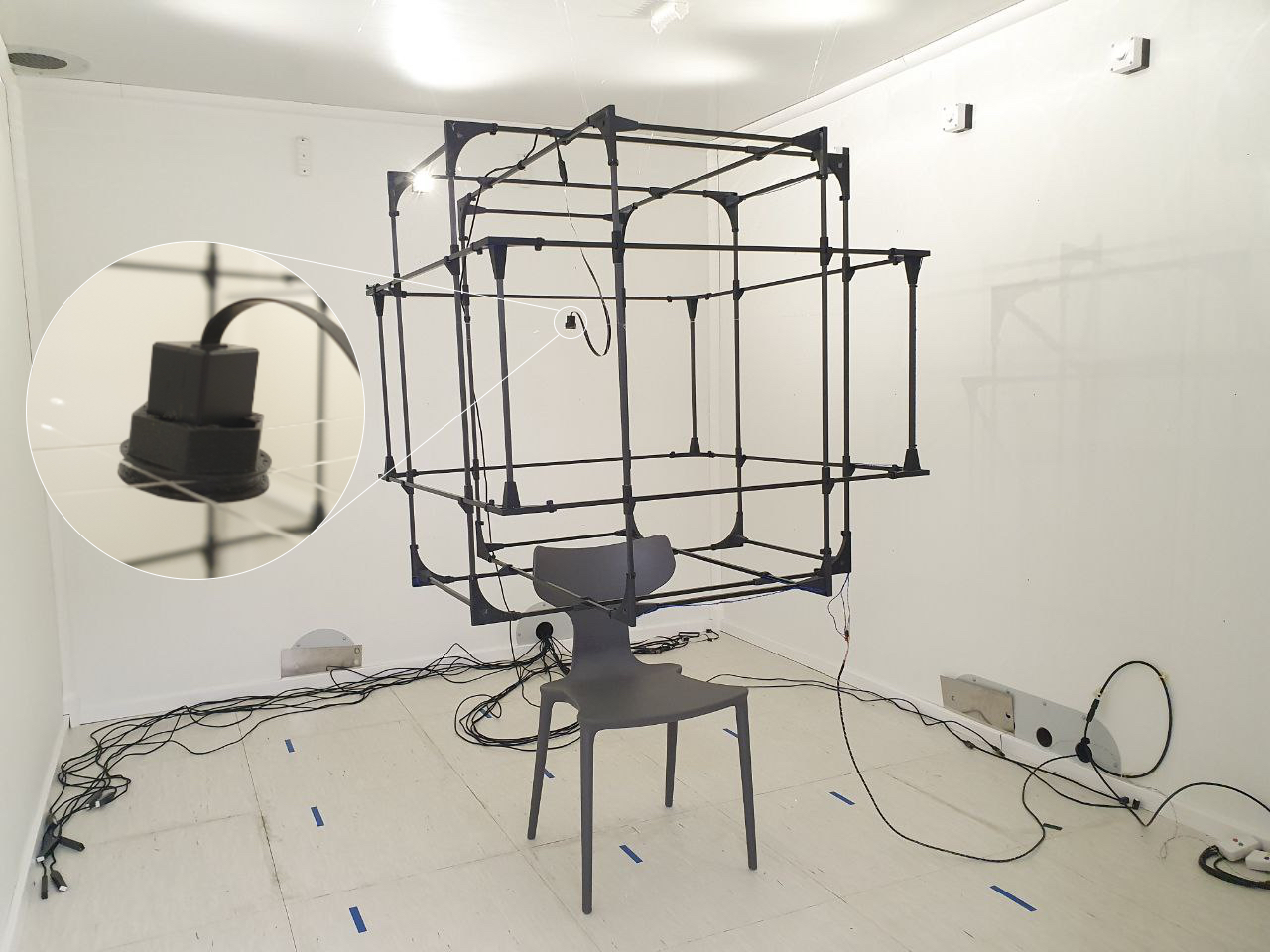} \\
		\end{tabular}
	\end{center}
	\caption{DASS's coils installed in room 2. Each pair of Helmholtz coils comprises square coils of  dimension $1~m \times 1~m$ located 0.5 $m$ apart and positioned in the three orthogonal planes. The QuSpin Gen-2 OPM, that is the internal reference sensor, is shown in the inset.}
	\label{fig:asas_photo}
\end{figure}

{The second environment where we repeated our experiments is provided by another Vacuumschmelze AK-3B MSR recently installed in the Higher School of Economics (HSE, Room 2) (www.hse.ru/en/cdm-centre/brain/). Here we implemented our own digital adaptive suppression system (DASS) capable of maintaining low ambient field within the sweet-spot. The diagram of the DASS is schematically drawn in Figure \ref{fig:asas}. It utilizes wire-loop coils, where the current is dynamically regulated  to compensate the local magnetic field fluctuations. The primary distinction of the DASS, compared to the analog system in Room 1, lies in its implementation of a two-reference scheme within a digital algorithm (least mean squares, LMS), which transforms the reference sensor signals into control signals for the coil drivers. The double-reference scheme in the LMS-based approach allows for the modeling of external noise (measured by an externally located, low-quality reference sensor) distortion caused by MSR penetration. Simultaneously, internal noise is measured by a single OPM-reference. This method enables the use of the external reference signal with a more precise frequency response compared to the analog system in Room 1. While the digital system offers flexibility and adjustability for remnant field suppression, it also carries the potential risk of additional noise being injection into the system, which we show to be negligible into the results section, however. The photography of the working DASS system is shown on the figure \ref{fig:asas_photo}.}

\subsection{Experimental paradigm, design and subjects}

{A total of 9 healthy volunteers participated in the experiments conducted across both MSR rooms. Five subjects completed the experiment in both rooms, three were new participants who were recorded exclusively in Room 2, and one subject took part solely in Room 1. During the experiment, subjects were seated comfortably and instructed to follow commands displayed on a screen. Motor tasks were employed to elicit sensorimotor rhythm (SMR) modulation in response to visual cues. Two tasks were used: a resting state, where participants were instructed to remain still and relaxed, and rhythmic grasping of the right hand, performed either overtly (for the motor execution paradigm, MR) or imagined (for the motor imagery paradigm, MI). Each condition lasted between 5 and 7 seconds, with random pauses of 1 to 2 seconds between conditions. The experimental timeline is depicted schematically in Figure 5, with a total of 30 randomly shuffled repetitions conducted for each condition. }

\begin{figure}
	\begin{center}
		\begin{tabular}{c}
			\includegraphics[width=0.85\linewidth]{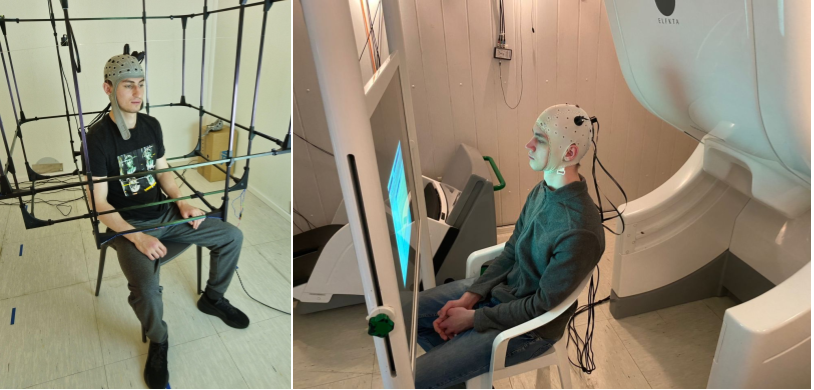} \\
		\end{tabular}
	\end{center}
	\caption{Participants in the two experimental environments: MEG Center MSR (right) and HSE MSR (left). The coil of the digital adaptive suppression system in the HSE room is also visible.}
	\label{subjects_sit}
\end{figure}

\begin{figure}
	\begin{center}
		\begin{tabular}{c}
			\includegraphics[width=0.85\linewidth]{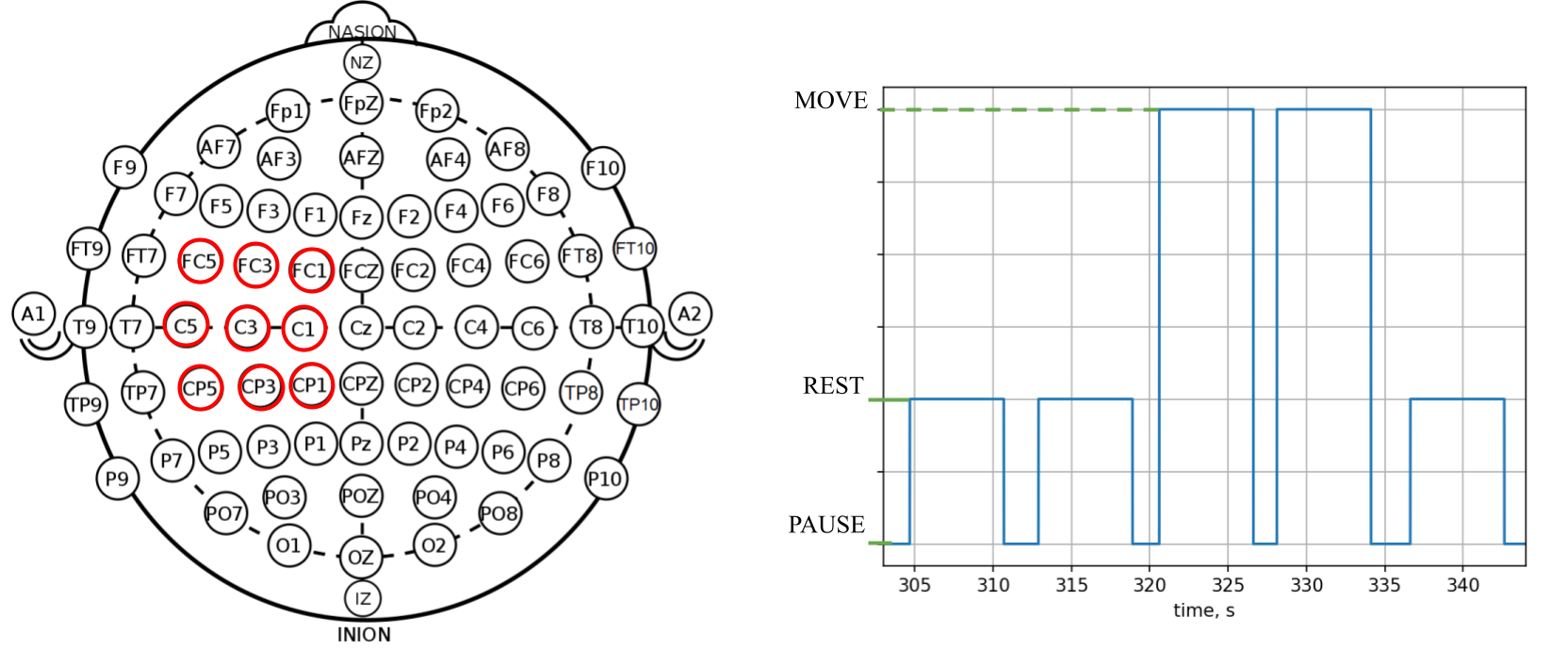} \\
		\end{tabular}
	\end{center}
	\caption{Sensor placement (left) based on the standard 10-10 system. A total of four sensors were positioned (with up to three reference sensors placed above them), and their exact locations varied between subjects. The positions consistently used are marked with red circles. The timeline snippet of the experiment (right) illustrates the pause, resting, and movement (or movement imagery) conditions.}
	\label{sensor_positioning}
\end{figure}

{For our purposes, we utilized commercially available zero-field sensors, including QuSpin Gen-2 and Gen-3 models (four Gen-2 sensors in Room 1 and a combination of seven Gen-2 and Gen-3 sensors in Room 2). These sensors were positioned close to the scalp using special holders attached to an elastic EEG cap. Data were recorded from a total of eight to fourteen channels, with half of the channels aligned to radial axes relative to the scalp, and the other half aligned to tangential axes. The precise orientation of the tangential channels was not rigorously controlled. Data acquisition was performed at a sampling rate of 1200 Hz. }

{The sensor locations were based on the 10-10 EEG montage positions \ref{sensor_positioning}. In Room 1, sensor placements varied across subjects as we sought to determine the optimal configuration for some participants. In Room 2 sensors were consistently positioned at C3, C1, and FC3. The locations were chosen to approximately cover the right-hand representation area in the sensorimotor cortex. In Room 1 no reference sensors were used for most subjects, except for Subject N1, where a gradiometer configuration was applied at the C3 location. In this setup, the reference sensor was positioned 7 cm above the active sensor in a special holder. In Room 2, due to the higher default background noise, gradiometers were employed at each position. Their construction followed a similar design, but the reference sensors were placed closer to the scalp, at a distance of 3.5 cm. }

\subsection{SMR Analysis}

{While the sensorimotor rhythm (SMR) can be observed and utilized for classification purposes (as discussed in the results section) even with individual sensors (without spatial filtering), our primary interest lies in the physiological sources of the SMR, which exhibit specific topographies and associations with motor behavior, distinct from artifacts (e.g., those resulting from subject movements) and external noise. To enhance our analysis, we applied the Common Spatial Pattern (CSP) filter as described by \cite{koles1990spatial} to the sensor data, yielding several components that maximized the contrast of source power in the resting state relative to the movement imagery (MI) and motor execution (MR) conditions. Prior to applying the CSP, we utilized a non-causal Butterworth third-order notch filter to reject power line noise (50 Hz and its harmonics) and bandpass the signal within the 7-33 Hz range ($\alpha$ + $\beta$  bands), effectively eliminating low-frequency non-physiological fluctuations and high-frequency technical noise.}

{We associated the extracted spatially distinct components with the physiological sensorimotor rhythm (SMR) and consider these components to be the primary building blocks for our analysis throughout the paper. We evaluated the spectral and spatio-temporal properties of the SMR component and its modulation by motor tasks. Initially, we calculated the power spectral densities (PSD) using the Welch method with a 1-second window, which provides a frequency resolution of 1 Hz. This analysis revealed $\alpha$ and $\beta$ spectral peaks as the argmax of the PSD within the predefined frequency bands (9-13 Hz for $\alpha$ and 15-25 Hz for $\beta$). We specifically focused on the $\mu$ and $\beta$ rhythms, as they are conventionally recognized as the most informative components of magnetoencephalography (MEG) signals for motor imagery brain-computer interface (MI-BCI) applications (\cite{pfurtscheller2006mu}), yielding consistent patterns of event-related desynchronization at the onset of MI and MR tasks. We quantified this pattern using a Mean Relative Desynchronization Metric \ref{MRD}.}

\begin{equation}\label{MRD}
	MRD = 10\cdot log_{10}(\frac{PSD_{rest}}{PSD_{move}})
\end{equation}

{In this equation, we extracted power spectral density (PSD) values within the 4 Hz and 6 Hz bands corresponding to the $\mu$ and $\beta$ components relative to their central frequencies. For most subjects, multiple spectral components were associated with the sensorimotor rhythm (SMR). To ensure systematicity, we selected only those spectral peaks with Mean Relative Desynchronization (MRD) values greater than 2.0, limiting our selection to no more than three peaks from three components for each band ($\alpha$ and $\beta$). If any component exhibited a pronounced peak in only one band (considered a weak component), we did not include it for the other band. All calculated values were subsequently compiled into a summary results table.}


\subsection{Pipeline for the offline-BCI experiment}

\begin{figure}
	\begin{center}
		\begin{tabular}{c}
			\includegraphics[width=0.95\linewidth]{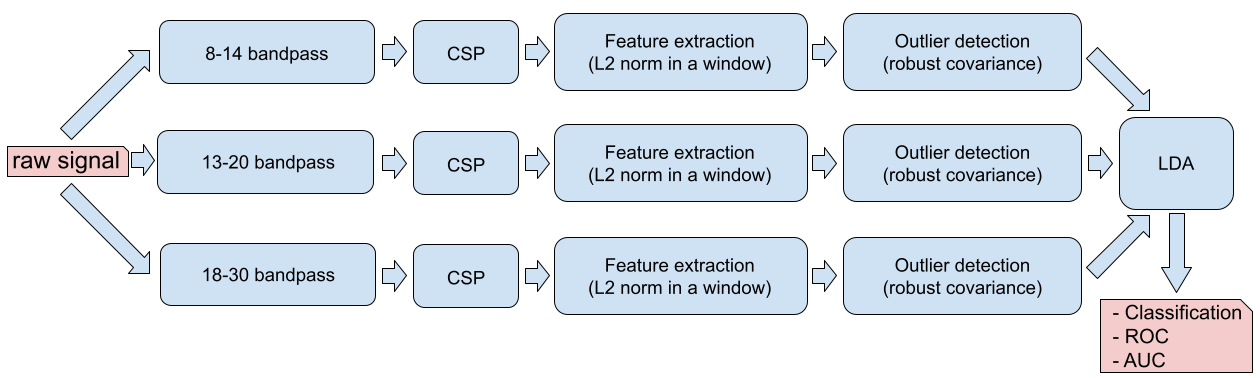} \\
		\end{tabular}
	\end{center}
	\caption{The pipeline of the offline-BCI. The raw signal is bandpassed in $\mu$, low-$\beta$, high-$\beta$ frequency ranges and then separately processed by the CSP-transformation, followed by the feature extraction and outlier dropping before the LDA-classifier}
	\label{csp_offline_bci}
\end{figure}

{The offline BCI analysis serves two primary purposes: to demonstrate the true informational content of the recorded signals and to assess their applicability to practical tasks such as motor imagery brain-computer interfaces (MI-BCI). While spectral analysis informs us about general features of separate SMR components, BCI performance metrics reveal how these components interact and how the activity is organized over time. We followed the classical Common Spatial Pattern (CSP) and Linear Discriminant Analysis (LDA) pipeline for linear classification, as illustrated in Figure \ref{csp_offline_bci}.}

 {The primary processing was similar to the CSP decomposition described in the SMR analysis section, with the exception that we divided the raw signal into three frequency bands: $\mu$ (8-14 Hz), low-$\beta$ (13-20 Hz), and high-$\beta$ (18-30 Hz). CSP was then applied separately to each band. Following this, feature extraction and outlier detection were performed using the robust covariance method (implemented via the sklearn Python package). We adopted a cautious approach, unlike many BCI studies, by selecting only components that exhibited higher synchronization during the motor task, thereby minimizing any artifact contamination in the features. While certain patterns, such as $\gamma$-activity, might display an inverse relationship, they are beyond the scope of this study. We set a threshold of 0.6 for CSP eigenvalues and selected all components exceeding this threshold, ensuring at least one component was retained per frequency band. Feature extraction was performed by calculating the $L_2$-norm for each CSP component within a predefined time window. While different time windows were tested, we standardized the duration to 1 second. For each subject, 10\% of the features were excluded. BCI performance was assessed using the Area Under the Curve (AUC) of the Receiver Operating Characteristic (ROC) for the binary classification task. AUC was calculated using a 5-fold cross-validation approach, where continuous data segments were cropped for testing, beginning at the start of the recording, with the crop shifted by the length of the testing segment for each iteration.}

\subsection{Real-time BCI}

\begin{figure}
	\begin{center}
		\begin{tabular}{c}
			\includegraphics[width=0.95\linewidth]{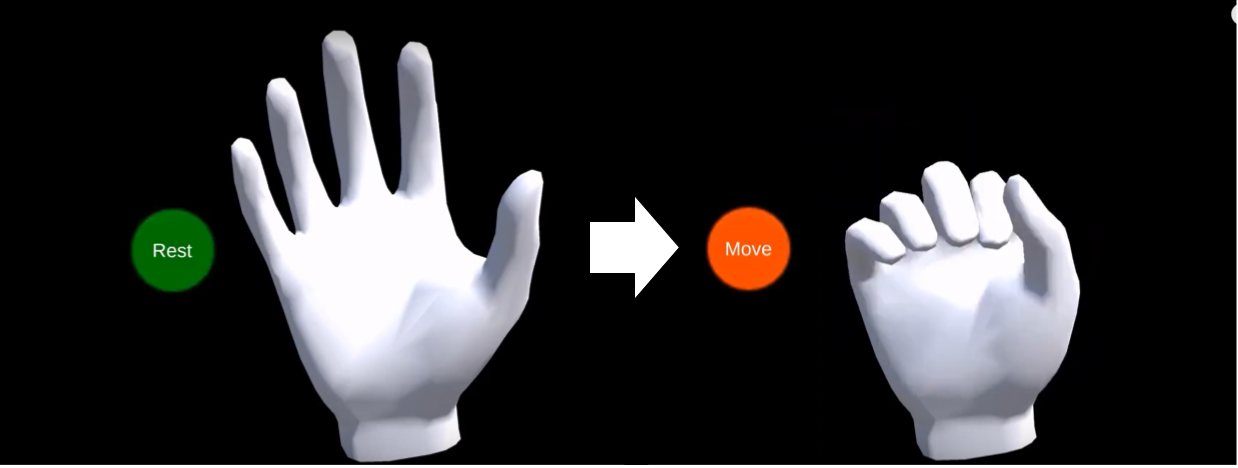} \\
		\end{tabular}
	\end{center}
	\caption{Phantom hand controlled by the subject during the online BCI experiment.}
	\label{phantom_hand}
\end{figure}

{The real-time BCI paradigm has distinct signal processing requirements that set it apart from the offline paradigm. First, all algorithms must operate in real time, necessitating the use of causal filters. Second, a balance must be achieved between the real-time control compatibility for the subject and the quality of the decoding. We reached this balance by achieving a decoding latency of approximately 1 second, coupled with smooth robust classification. The experiment was divided into two parts: a training part, where processing parameters were quickly adjusted, and a testing part, in which the subject attempted to control a virtual hand via motor imagery (MI) in response to alternating on-screen commands (Figure \ref{phantom_hand}). The data interfacing was managed through MATLAB, with data processing and command generation performed in Python, while the two-state virtual hand (rhythmic hand grasp-release or fixed open hand) was implemented in Unity.}

\section{Results}

\subsection{DASS provides favorable environment tor the OPM-experiments}

\begin{figure}[H]
\centering
\begin{subfigure}[t]{.48\textwidth}
    \centering
    \includegraphics[width=\textwidth]{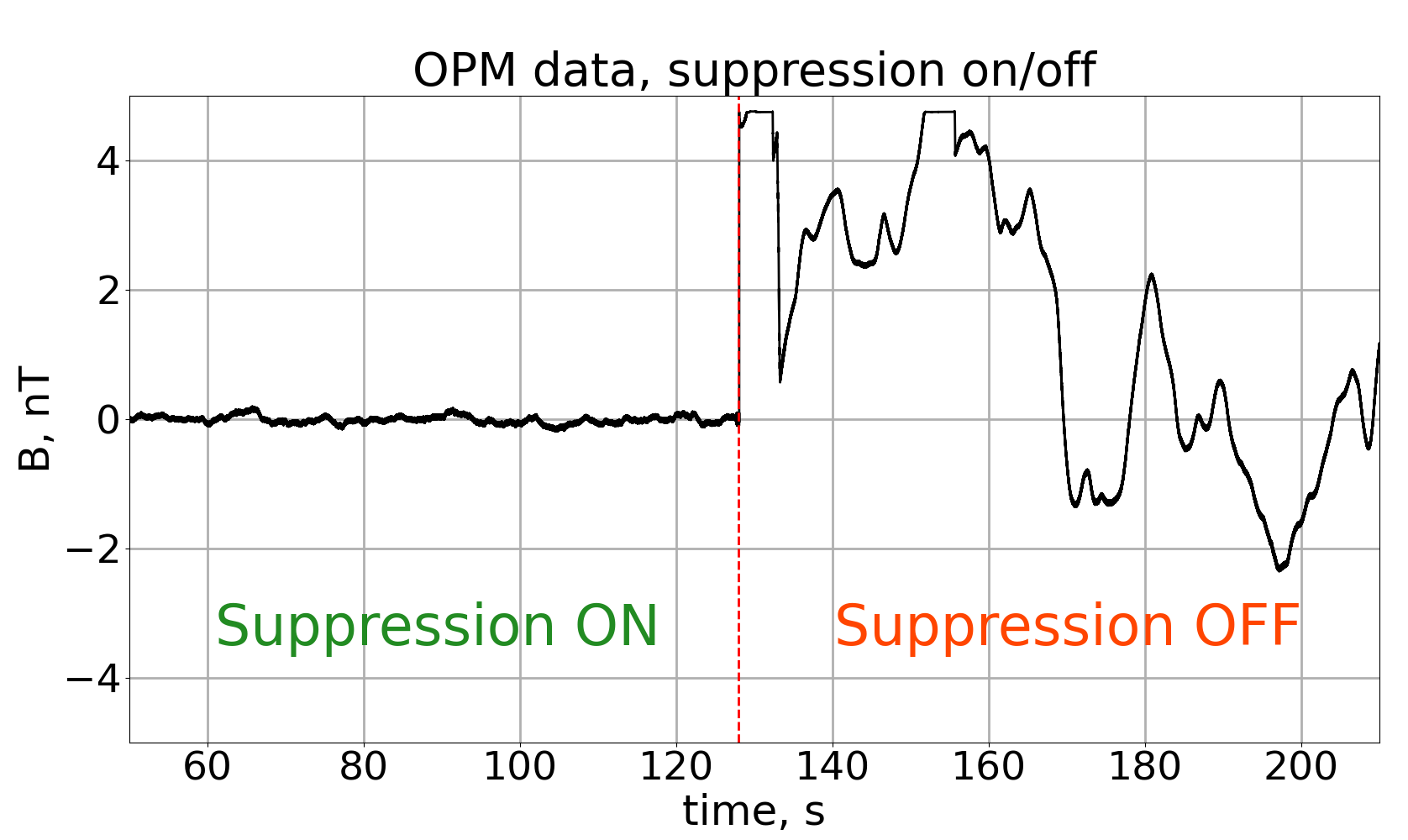}
    \caption{}
    \label{fig:curves} 
\end{subfigure}
\begin{subfigure}[t]{.48\textwidth}
   \centering
   \includegraphics[width=\textwidth]{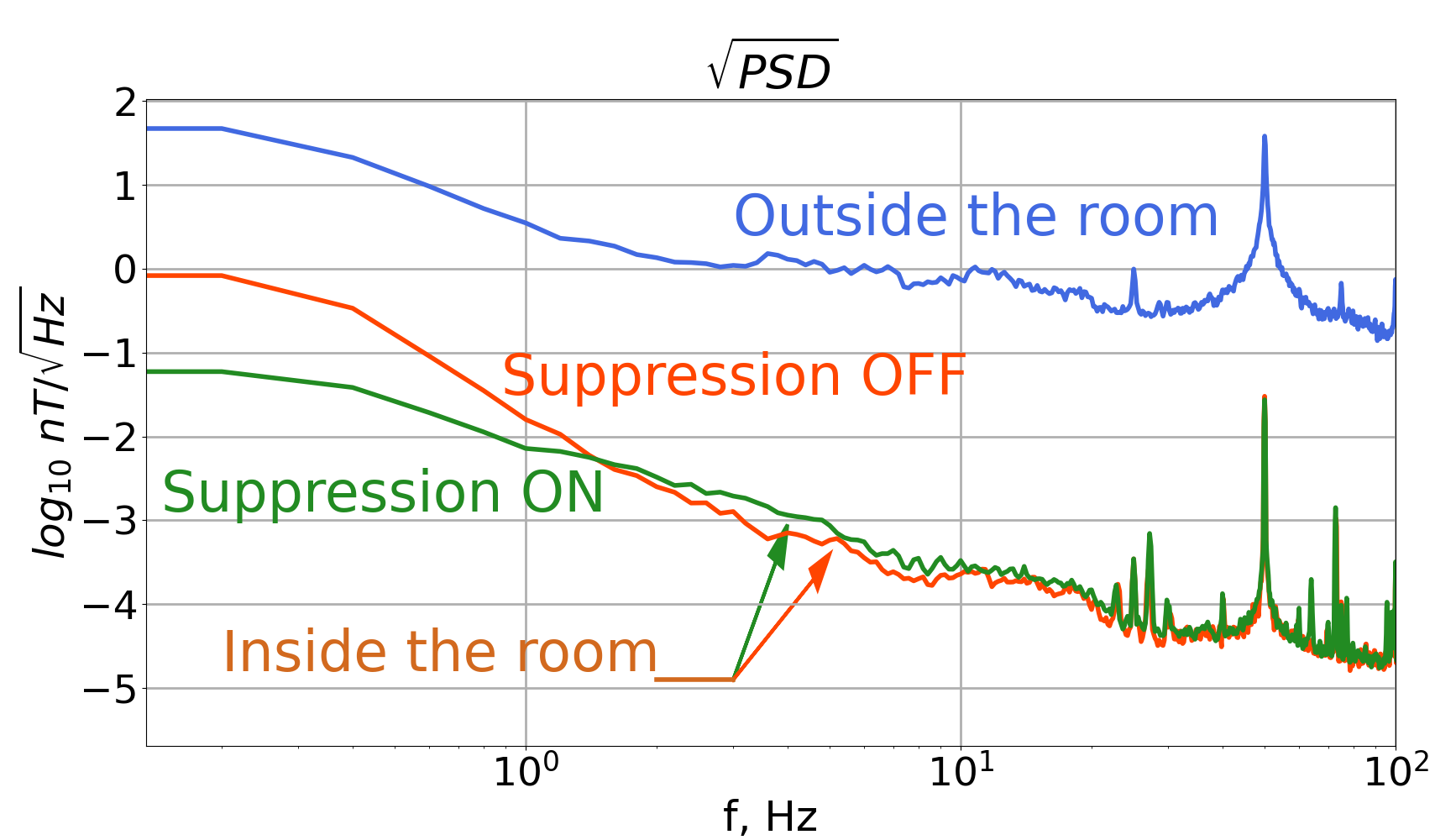}
   \caption{}
   \label{fig:msr}
\end{subfigure}
\caption{a) The QuSpin sensor signal during the DASS OFF period and DASS on period. b) PSD of the background noises recorded outside the MSR (room 2), inside the room with DASS OFF, and inside the ROOM with DASS ON.}
\label{DASS_perfomance}
\end{figure}

{Figure \ref{DASS_perfomance} illustrates the performance of the DASS system in room 2. Residual external magnetic field induce high-range fluctuations in the OPM signal, with saturation visible in certain areas of the plot (a) when DASS is inactive. When DASS is engaged, the signal stabilizes, remaining within the optimal zero-field range. The PSD plots in (b) reveal limitations of the MSR room: at ultralow frequencies (<1 Hz), the shielding factor (the distance between the Outside the room and Inside the room, suppression OFF curves) decreases significantly, by up to $10^2$ times compared to higher frequencies. DASS mitigates this distortion and enhances the shielding effect (suppression ON curve Inside the room). Although the DASS system significantly improves signal quality, it introduces a minor noise artifact within the 10-20 Hz band, arising from various stages of signal processing inside DASS. However, this artifact is consistent and can be effectively eliminated through spatial filtering, ensuring minimal impact on SMR signal quality. An enhanced version of DASS is currently under development. Overall, the DASS system provides a highly suitable environment for QuSpin OPM sensors, enabling accurate SMR rhythm recordings.}

\subsection{Mean Relative Desynchronization of SMR-rhythm magnetic field}

{
Table \ref{MRD MI} presents the MRD values and central frequencies for each participant and all SMR components (selected as described in the methods section) during the imaginary movement session, while Table \ref{MRD MR} shows the same data for the real movement sessions. The calculated MRD values are the primary characteristics of the individual SMR-rhythm components, as they reflect both the peak sharpness of the rhythm in the resting state and its reactivity to motor task performance. However, they can be sensitive to the subject's experience and quality of the his performance during contrastive tasks, particularly in MI scenarios, where the task is processed covertly. In line with this expectations, the values in the table show that most participants exhibited components with expressive desynchronization behavior in both MI and MR paradigms, usually with slightly lower values in MI comparing to MR. }

\begin{figure}
	\begin{center}
		\begin{tabular}{c}
			\includegraphics[width=0.98\linewidth]{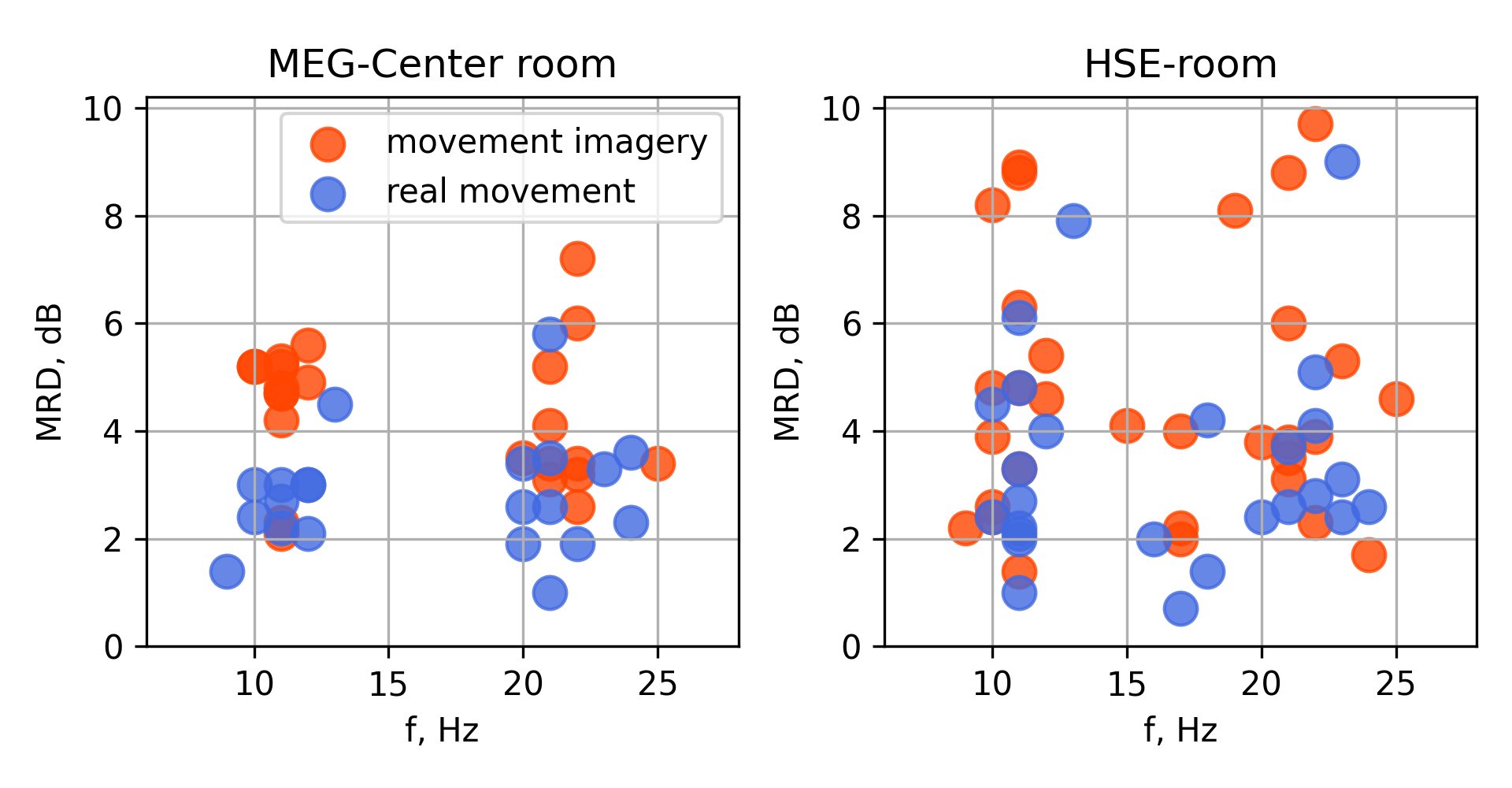} \\
		\end{tabular}
	\end{center}
	\caption{The scatter plot of the MRD values and peak frequencies for room 1 and room 2, MI paradigm (blue bubbles), and MR paradigm (orange bubbles).}
	\label{summary_table}
\end{figure}

{The tables were also visualized in \ref{summary_table}, allowing for clear comparisons of MRD patterns across different conditions. The visualization shows that MRD values varied significantly across participants, although most points are compactly clustered within the 2 dB to 6 dB MRD range, and between 10 Hz to 12 Hz for the $\alpha$ band and 20 Hz to 23 Hz for the $\beta$ band. From the table and visualization, it is evident that within each subject, MRD values for the $\alpha$ and $\beta$ bands are similar, reflecting the intrinsic connection between these rhythms as part of a unified sensorimotor rhythm.}

\begin{table}
	\caption{Central frequencies, MRD values, AUC-metrics for imagery movements. Each next component starts from a new line}
	\centering
	\begin{tabular}{|c|p{4.5cm}|p{1.5cm}|p{4.5cm}|p{1.5cm}|}
		\toprule
		Subject     &  room 1 \newline central freq (MRD) & room 1 \newline AUC &room 2 \newline central freq (MRD) & room 2 \newline AUC \\ 
		\midrule
		Subj. N1 & 12Hz (3.7dB), 21Hz (5.8dB)\newline 12Hz (3.0dB), 21Hz (3.5dB) \newline 21Hz (2.6 dB) & 0.97  & 11Hz (2.2dB), 24Hz (2.6dB) &  0.87   \\
        \hline
		Subj. N2  & 12Hz (3.0dB), 24Hz (3.6dB) \newline 11Hz (2.7dB), 23Hz (3.3dB) \newline 12Hz (2.1dB), 24Hz (2.3dB)& 0.76 & 11Hz (1.0dB), 18Hz (1.4dB) & 0.66      \\
        \hline
        Subj. N3   & 11Hz (2.2dB) \newline 20Hz (1.9dB)   & 0.73    & 12Hz (4.0dB) 23Hz (2.4dB) & 0.81 \\
        \hline
        Subj. N4   & 10Hz (3.0dB), 20Hz (2.6dB)\newline 10Hz (2.4dB), 20Hz (3.4dB)  & 0.76 & 10Hz (4.5dB) 21Hz (3.7dB) \newline  10Hz (2.4dB) 21Hz (2.6dB) \newline & 0.84\\
        \hline
        Subj. N5   & 9Hz (1.4dB), 22Hz (1.9dB)    & 0.65   & 11Hz (2.0dB)\newline 11Hz (2.1dB) \newline 17Hz (0.7dB) & 0.79 \\
        \hline
        Subj. N6     & NOT RECORDED & N. R.      & 11Hz (4.8dB) 22Hz (4.1dB) \newline 20Hz (2.4dB)  & 0.87 \\
        \hline
        Subj. N7    & NOT RECORDED  & N. R.     & 11Hz (6.1dB), 22Hz (5.1dB)\newline 11Hz (3.3dB), 22Hz (2.8dB) & 0.87  \\
        \hline
        Subj. N8   & NOT RECORDED & N. R.       & 13Hz (7.9dB), 18Hz (4.2dB), 23Hz (9.0 dB)\newline 11Hz (2.7 dB), 23Hz (3.1 dB)\newline 16Hz (2.0dB) &0.89 \\
        \hline
        Subj. N9    & 11Hz (3.0 dB) 21Hz (1.0 dB)  & 0.80 & NOT RECORDED & N. R.  \\
		\bottomrule
	\end{tabular}
	\label{MRD MI}
\end{table}

\begin{table}
	\caption{Central frequencies, MRD values for real movements. Each next component starts from a new line}
	\centering
	\begin{tabular}{|c|p{6.4cm}|p{6.4cm}|}
		\toprule
		Subject     &  room 1 \newline central freq (MRD)  &room 2 \newline central freq (MRD)\\ 
		\midrule
		Subj. N1 & 13Hz (4.5 dB), 25Hz (3.4 dB)\newline 12Hz (4.9 dB), 21Hz (3.1 dB)   & 12Hz (4.6dB), 24Hz (1.7dB)     \\
        \hline
		Subj. N2  & 11Hz (5.2dB), 22Hz  (7.2dB)\newline 11Hz (4.7dB), 22Hz (6.0dB)\newline 22Hz (3.4dB) & 11Hz (1.4dB), 17Hz (4.0dB)  \newline 25Hz (4.6dB)   \\
        \hline
        Subj. N3   & 11Hz (4.8dB), 22Hz (3.2dB)\newline 11Hz (2.3dB)      & 11Hz (8.8dB), 23Hz (5.3dB)\newline 12Hz (5.4dB) \\
        \hline
        Subj. N4   & 10Hz (5.2dB), 21Hz (5.2dB)\newline 10Hz (5.2dB), 20Hz (3.5dB)   \newline 21Hz (3.4dB)   & 10Hz (8.2dB), 21Hz (8.8dB)\newline 10Hz (3.9dB), 21Hz (3.8dB) \newline 10Hz (2.6dB), 21Hz (3.1dB) \\
        \hline
        Subj. N5  & 11Hz (4.2dB) 22Hz (2.6dB)  & 10Hz (4.8dB), 20Hz (3.8dB) \newline 10Hz (2.4dB), 17Hz (2.0dB) \newline 11Hz (4.8dB)     \\
        \hline
        Subj. N6     & NOT RECORDED       & 11Hz (8.9dB), 21Hz (6.0dB)\newline 21Hz (3.5dB)  \\
        \hline
        Subj. N7     & NOT RECORDED     &  11Hz (3.3), 22Hz (3.9)  \\
        \hline
        Subj. N8   & NOT RECORDED       & 11Hz (6.3dB), 19Hz (8.1dB), 22Hz (9.7 dB)\newline 9Hz (2.2dB), 15Hz (4.1dB)\newline17Hz (2.2dB), 22Hz (2.3dB)  \\
        \hline
        Subj. N9     & 11Hz (5.3dB), 21Hz (4.1dB) \newline 11Hz (4.7dB) \newline 11Hz (2.1 dB)     & NOT RECORDED  \\
		\bottomrule
	\end{tabular}
	\label{MRD MR}
\end{table}

\begin{figure}
	\begin{center}
		\begin{tabular}{c}
			\includegraphics[width=0.95\linewidth]{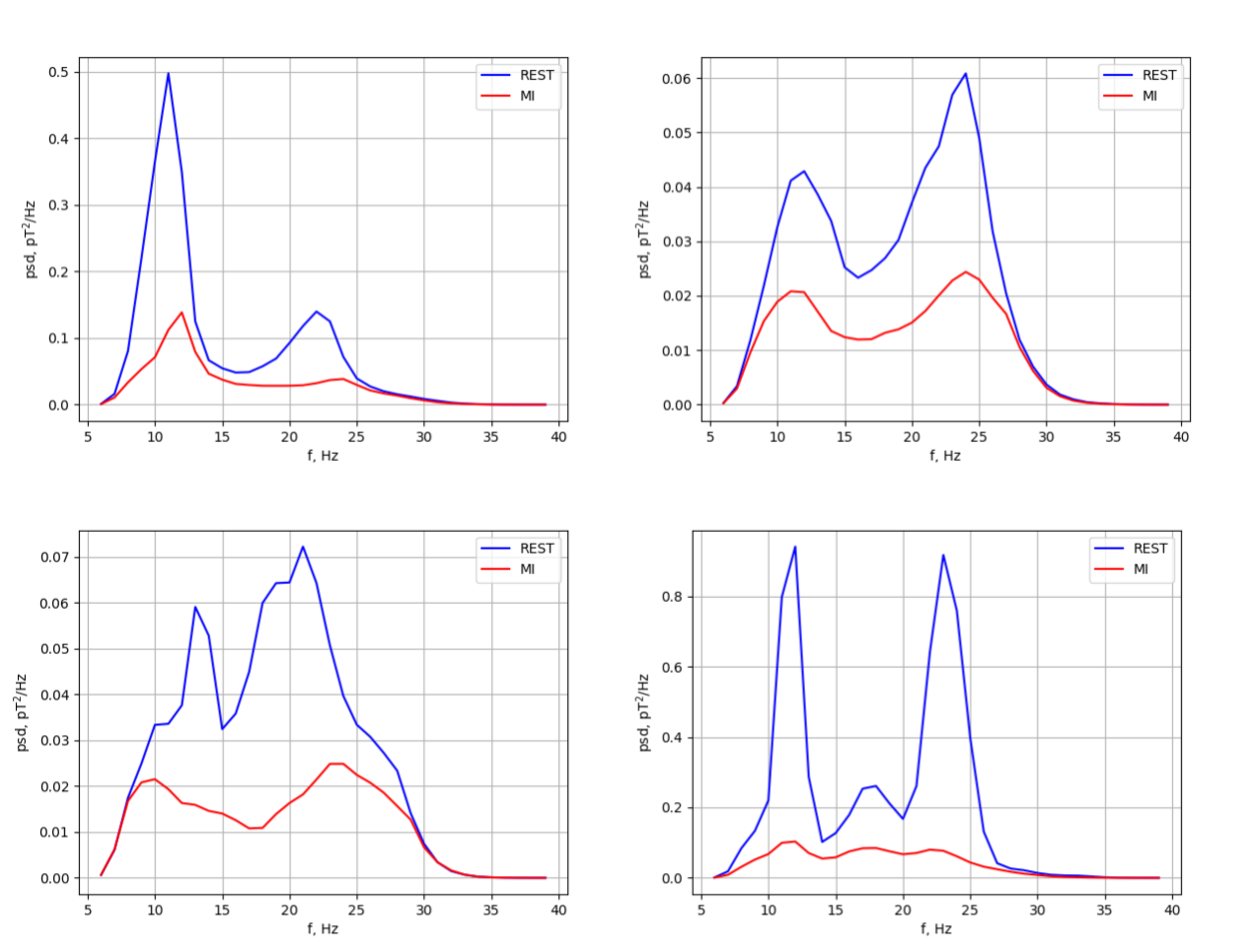} \\
		\end{tabular}
	\end{center}
	\caption{The diversity of SMR-components for different subjects. They can be with prevalent $\alpha$-peak, $\beta$-peak, with both of them and even with three peaks. Notably, all the SMR components presented here are for MI paradigm.}
	\label{psds}
\end{figure}

{Most of the spatiotemporal components exhibited the stereotypical form of the SMR rhythm with two peaks ($\alpha$ and $\beta$), though some components were dominated by either the $\mu$ or $\beta$ rhythms. There was even a component with three peaks ($\alpha$, low $\beta$, and high $\beta$) observed in one subject. Figure \ref{psds} illustrates the diversity of power spectral density profiles of SMR components across different subjects. The spectral power varied greatly between subjects and components, although for most components, the $\alpha$ peak power was dominant.}

\begin{figure}
\centering
\begin{subfigure}[t]{.48\textwidth}
    \centering
    \includegraphics[width=\textwidth,trim=0.5cm 0 1cm 0, clip]{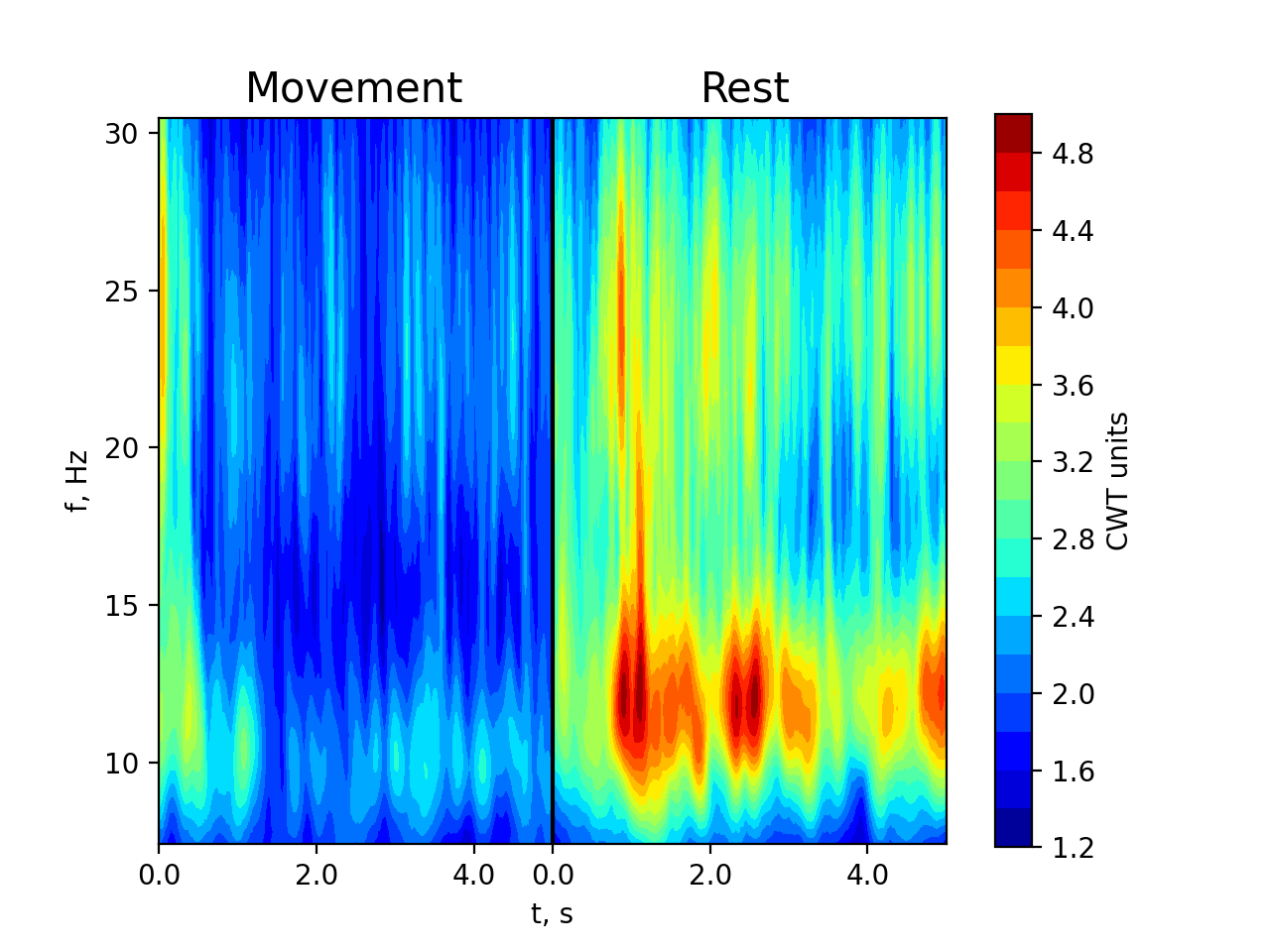}
    \caption{}
    \label{fig:curves} 
\end{subfigure}
\begin{subfigure}[t]{.48\textwidth}
   \centering
   \includegraphics[width=\textwidth,trim=0.5cm 0 1cm 0, clip]{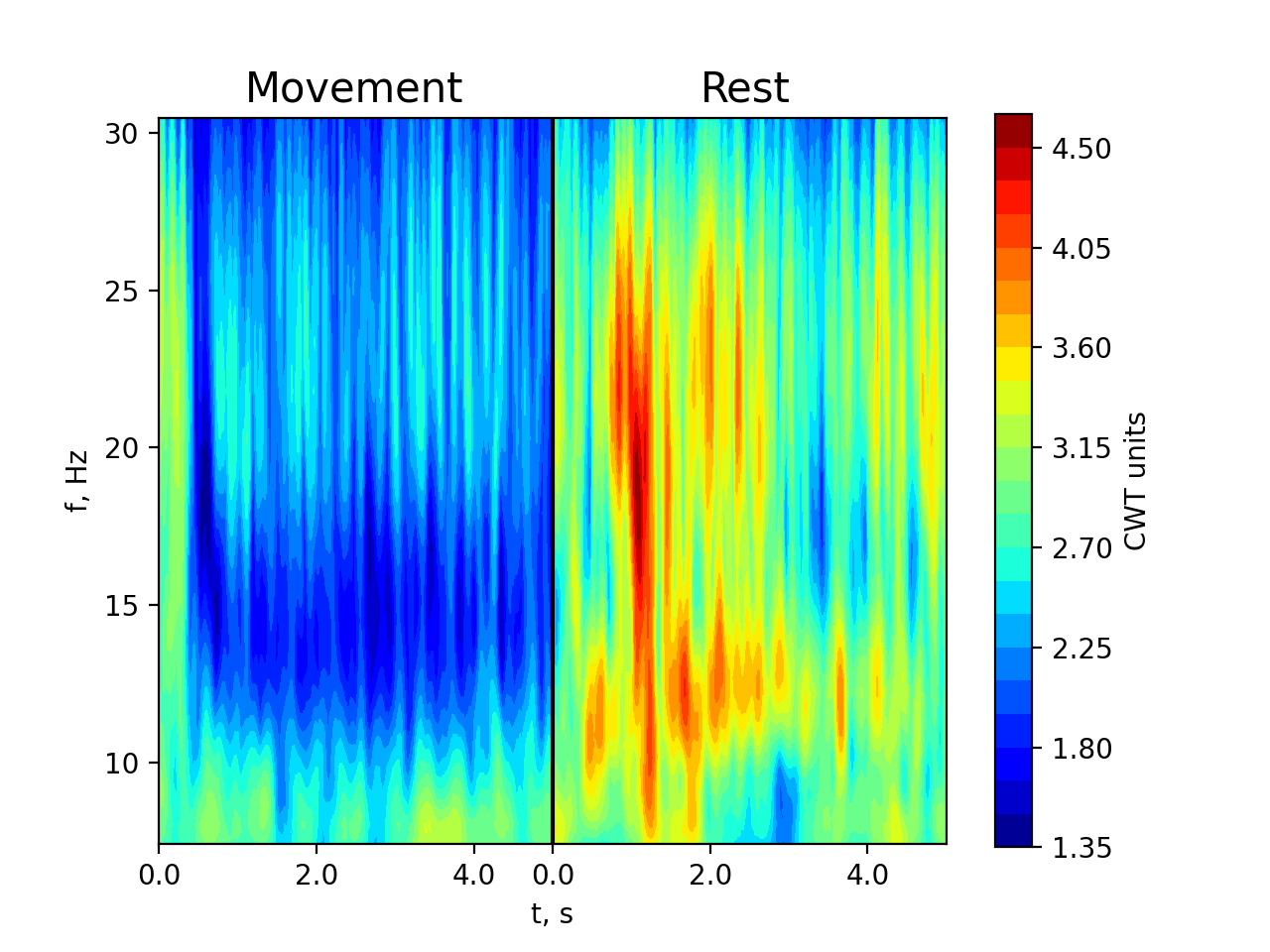}
   \caption{}
   \label{fig:msr}
\end{subfigure}
\begin{subfigure}[t]{.48\textwidth}
   \centering
   \includegraphics[width=\textwidth,trim=0.5cm 0 1cm 0, clip]{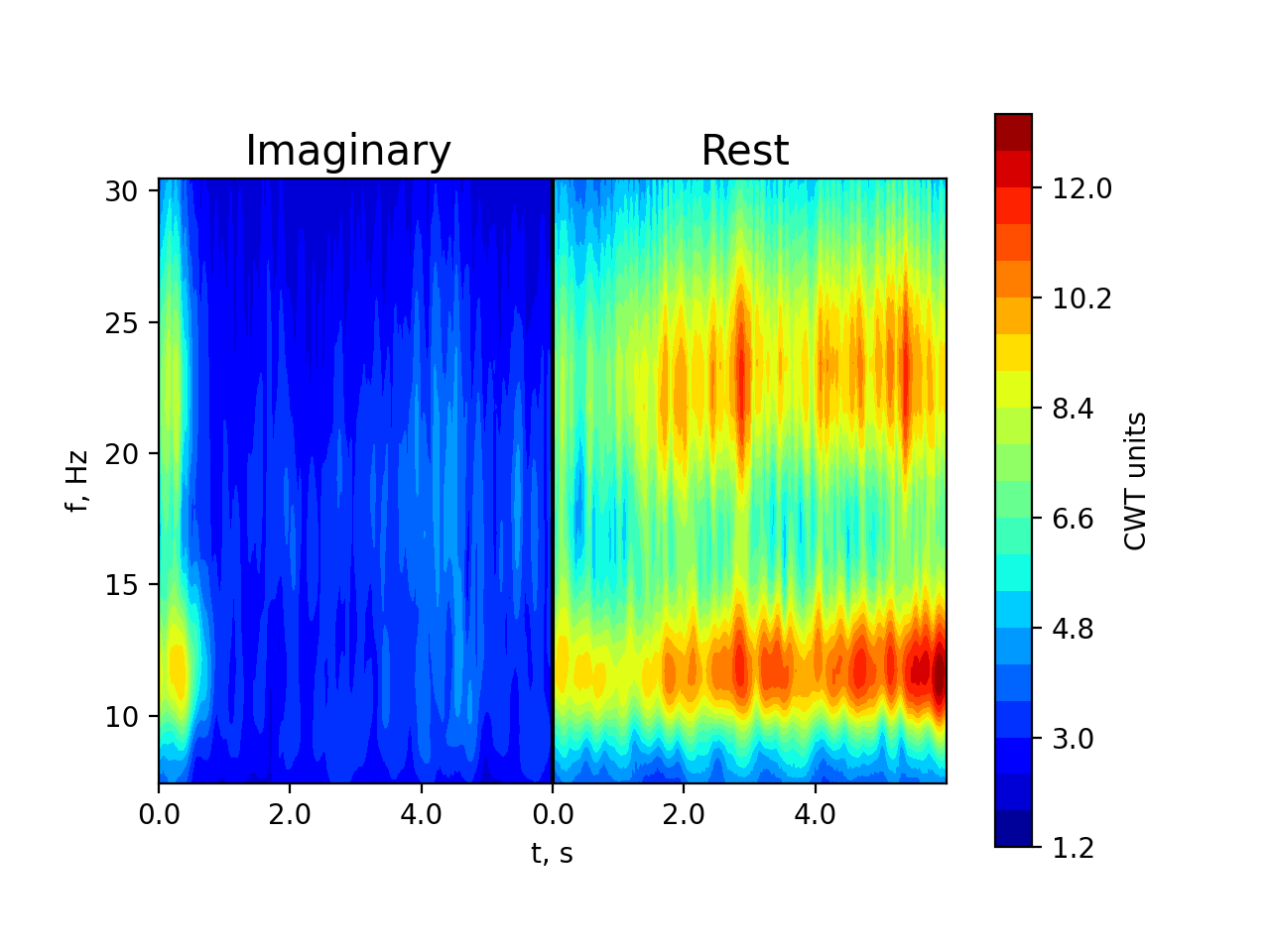}
   \caption{}
   \label{fig:msr}
\end{subfigure}
\begin{subfigure}[t]{.48\textwidth}
   \centering
   \includegraphics[width=\textwidth,trim=0.5cm 0 1cm 0, clip]{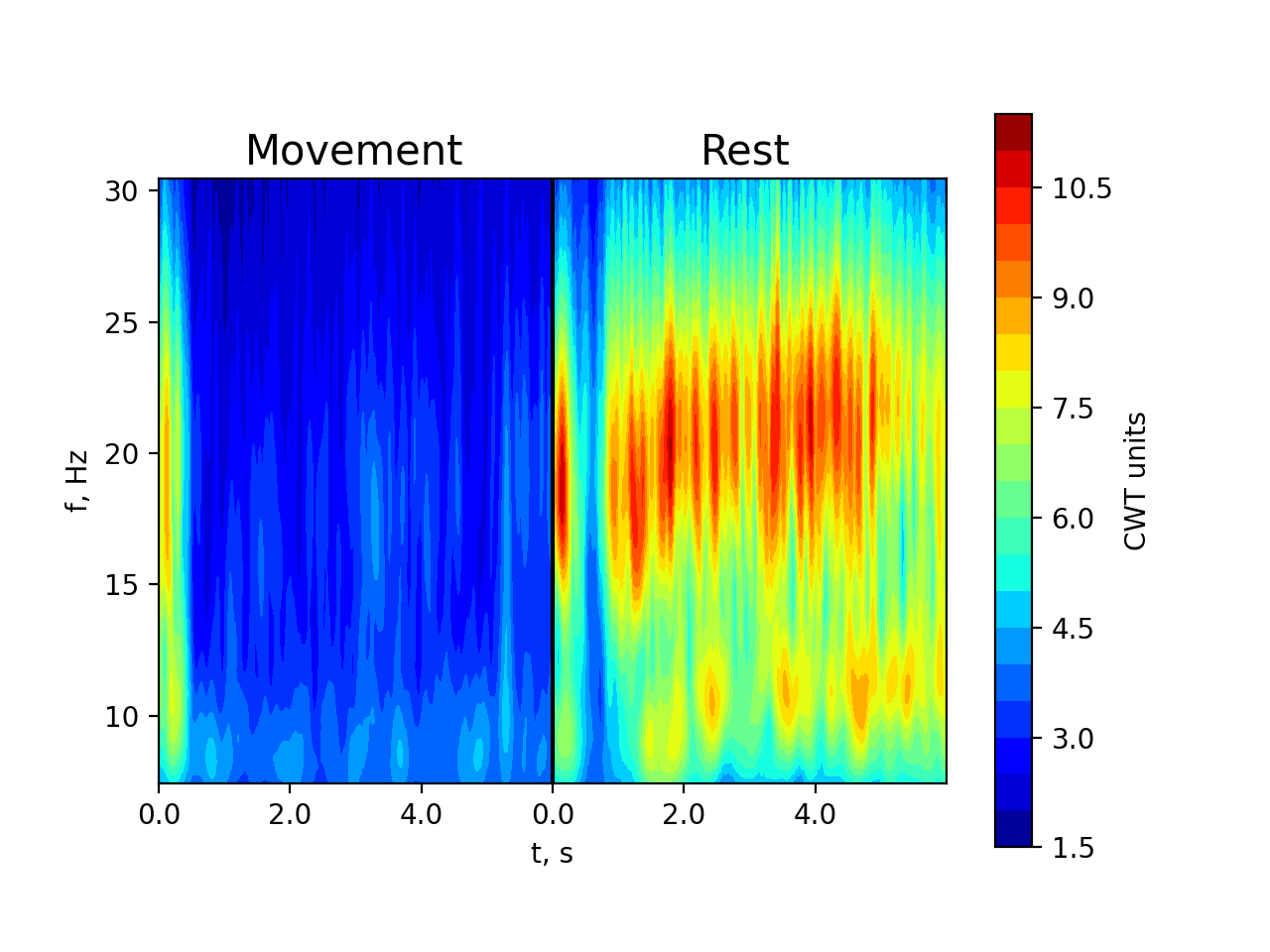}
   \caption{}
   \label{fig:msr}
\end{subfigure}
\begin{subfigure}[t]{.48\textwidth}
   \centering
   \includegraphics[width=\textwidth,trim=0.5cm 0 1cm 0, clip]{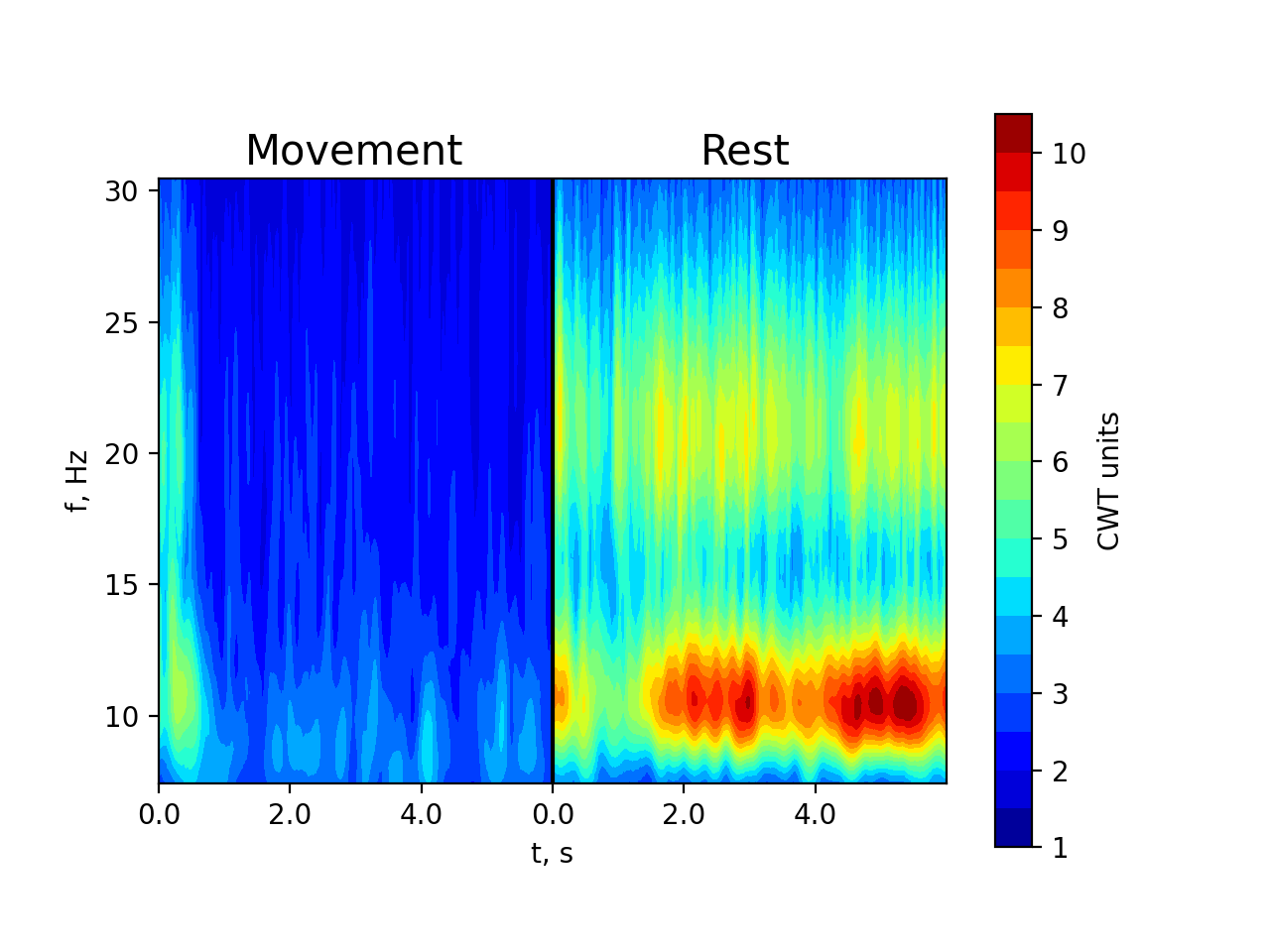}
   \caption{}
   \label{fig:msr}
\end{subfigure}
\begin{subfigure}[t]{.48\textwidth}
   \centering
   \includegraphics[width=\textwidth,trim=0.5cm 0 1cm 0, clip]{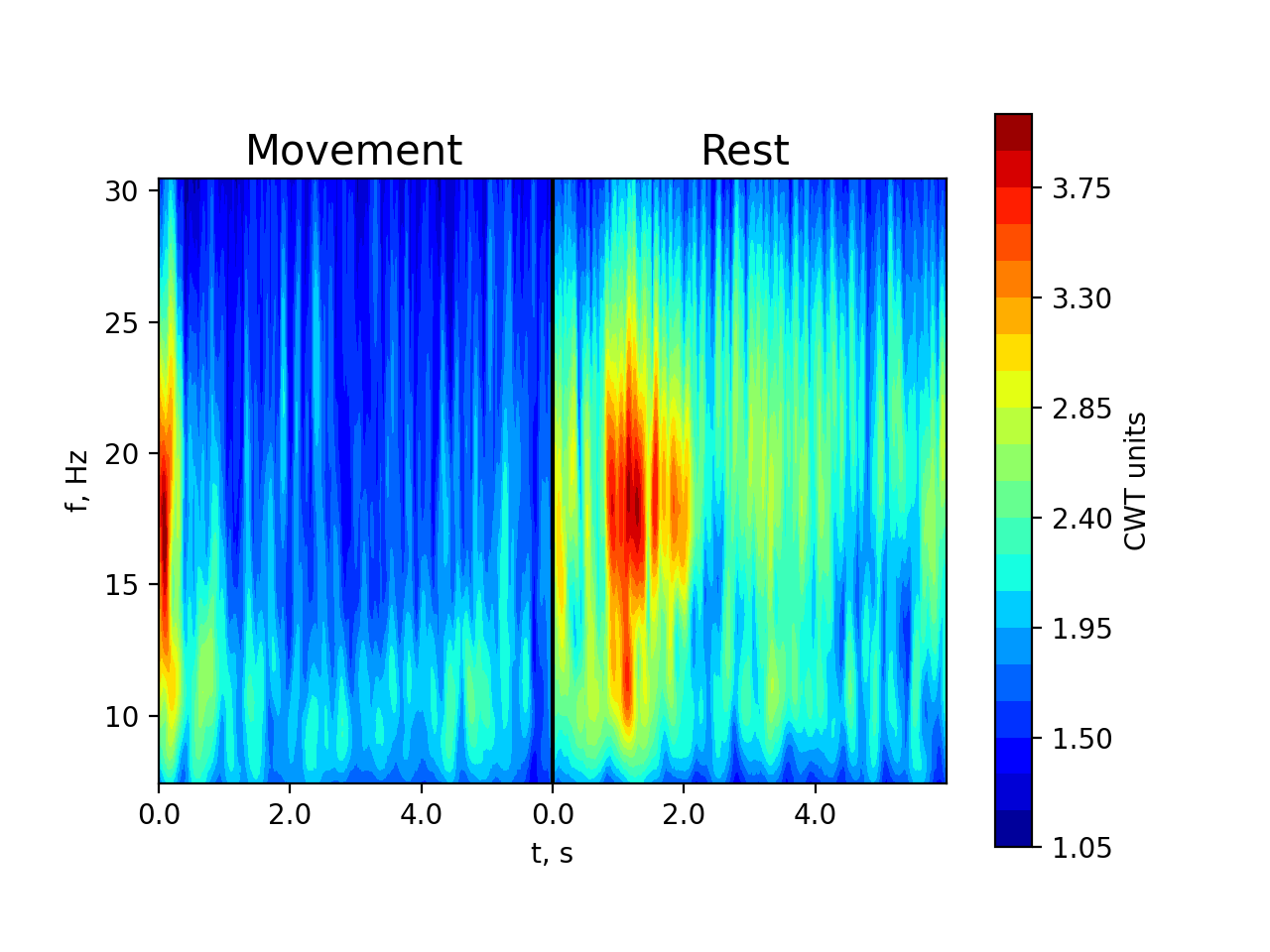}
   \caption{}
   \label{fig:msr}
\end{subfigure}
\caption{Averaged across epochs continuous wavelet transform envelope for different subjects in two conditions: MR/MI (left half) and REST (right half).}
\label{spectrograms}
\end{figure}

{
While recording from a relatively small area with a low count of OPM sensors, the spatial filter was able to identify at least two SMR components for most subjects in both MI and MR conditions. For some participants, the spatial and spectral forms were similar, potentially reflecting the broad spatial extent of SMR desynchronization. For others, however, the spectral forms varied (see \ref{spectrograms}, (a) and (b), showing two components from the same session), reflecting the high spatial resolution of the OPM measurements, which allows signals from closely located brain sources with different spectral properties to be captured.}

{
Other averaged time-frequency plots in \ref{spectrograms} demonstrate short, overlapping spindles of rhythmic activity from different epochs, with varying presence in the $\alpha$ and $\beta$ domains, as well as clear and broad desynchronization (especially in (c), (d), and (e)). A lag artifact at the beginning of the trials is observed, caused by task perception latency. For some participants, a clear $\beta$ rebound effect was observed. Spectrograms (b) and (f) show a brief increase in power in the $\beta$ and $\alpha$ domains at the start of resting trials. We note that this effect might have been more pronounced if all resting trials had directly followed motor-task trials. However, in our study, resting trials occurred after a short pause, so the rebound effect is likely only visible for participants who was concentrating during the pause.}

\subsection{Offline BCI}

{We assessed the performance of the offline BCI paradigm for MI sessions using AUC metrics, with results summarized in Table \ref{MRD MI}. BCI performance is influenced by factors including rhythm prominence, spectral definition, subject engagement, sensor placement, and individual variability. In all sessions, BCI performance exceeded chance level, with one session in room 1 and four sessions in room 2 showing exceptional performance, surpassing an AUC value of 0.85.}

{For one subject in room 1 and two subjects in room 2 with optimal performance, we investigated the impact of sensor quantity and time window (including reference sensors) on data informativeness \textcolor{red}{\ref{}}. The smooth gradient of AUC values in figure indicates that successful classification requires high SNR, achievable through either multichannel recordings or extended time windows for signal accumulation. In each of the three cases, a single OPM sensor placed at an informative location on the head proved sufficient to achieve decoding quality comparable to that obtained with all sensors.}

\begin{figure}
\centering
\begin{subfigure}[t]{.318\textwidth}
    \centering
    \includegraphics[width=\textwidth,trim=1cm 0 3cm 0, clip]{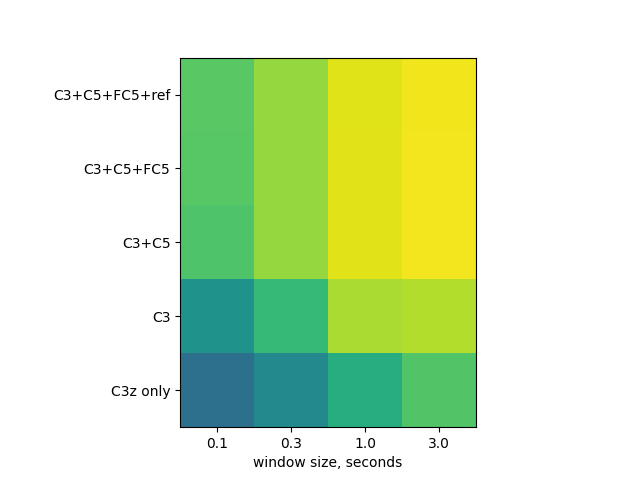}
    \caption{}
    \label{fig:curves} 
\end{subfigure}
\begin{subfigure}[t]{.318\textwidth}
   \centering
   \includegraphics[width=\textwidth,trim=1cm 0 3cm 0, clip]{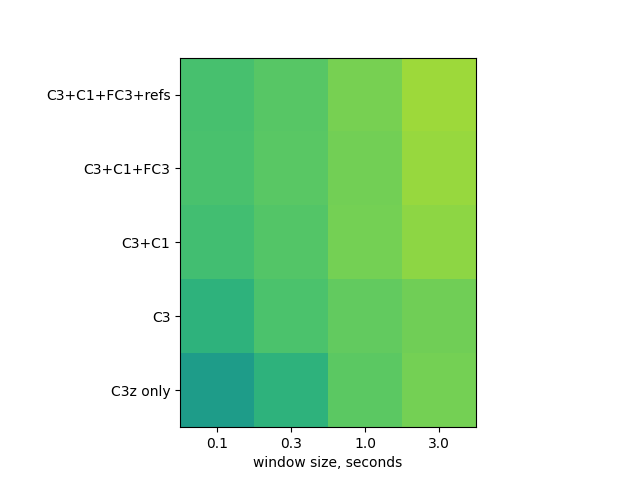}
   \caption{}
   \label{fig:msr}
\end{subfigure}
\begin{subfigure}[t]{.353\textwidth}
   \centering
   \includegraphics[width=\textwidth,trim=1cm 0 1.5cm 0, clip]{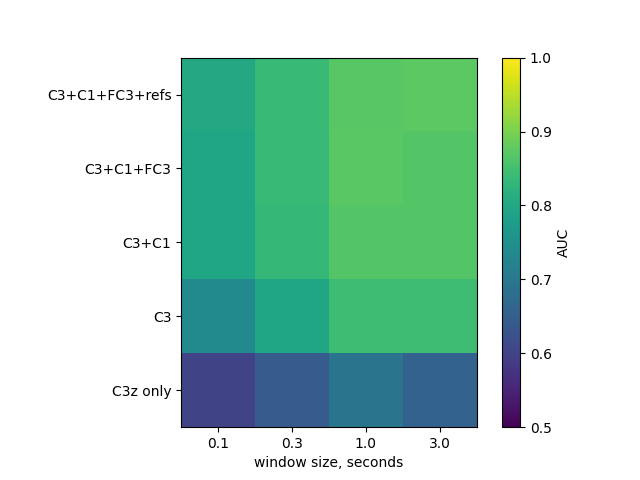}
   \caption{}
   \label{fig:msr}
\end{subfigure}
\caption{AUC values (room 1, subj. N1, room 2, subj. N8, room 2, subj. N7) depending on the number of sensors and feature extraction time-window used.}
\label{gradient}
\end{figure}

\begin{figure}
	\begin{center}
		\begin{tabular}{c}
			\includegraphics[width=0.5\linewidth]{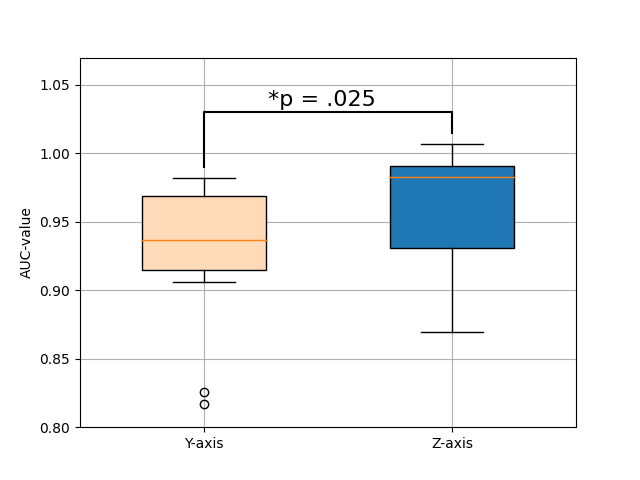} \\
		\end{tabular}
	\end{center}
	\caption{The boxplot of the difference between Z-axis-based classification AUC values and Y-axis-based.}
	\label{boxplot}
\end{figure}

{We also investigated the informativeness of different measurement axes of the OPM sensor—specifically, the radial (Z) axis oriented perpendicular to the scalp and the tangential (Y) axis. To accomplish this, we extracted features separately from each axis and normalized the resulting AUC values against those from the full set of channels, subsequently testing for differences \ref{boxplot}. A paired Wilcoxon test revealed a statistically significant advantage (p = 0.025) of the radial axis field over the tangential axis field for the MI task classification.}

\subsection{Real-time BCI}

\begin{figure}
	\begin{center}
		\begin{tabular}{c}
			\includegraphics[width=1.0\linewidth]{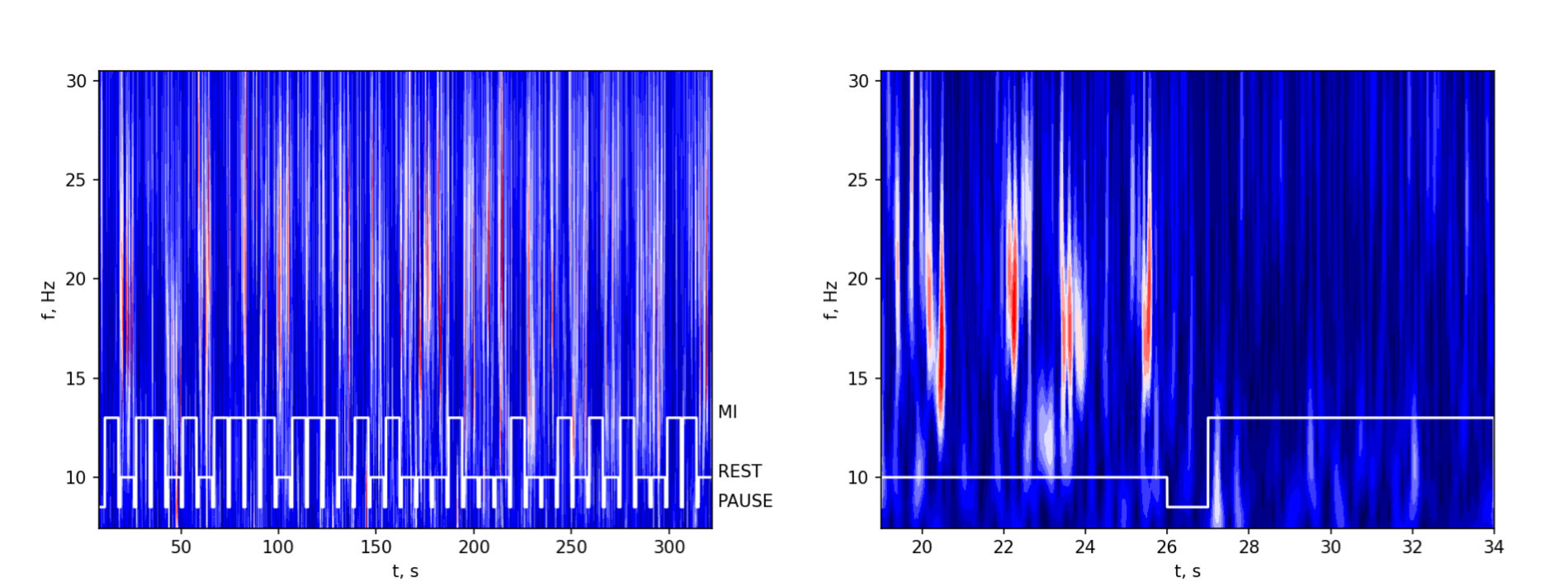} \\
		\end{tabular}
	\end{center}
	\caption{The time-frequency plot for the real-time BCI session. The white level-line illustrates the tasks subject had to perform at the moment}.
	\label{unfolded_spctr}
\end{figure}

{Finally, we evaluated our BCI in real-time, with the subject controlling a virtual hand (see Methods section) that could alternate between active and resting states based solely on the subject’s imagined intentions. Several tests were conducted, achieving optimal control latency and smooth, reliable decoding. The operator (subj. N1) reported a clear sense of direct control over the virtual hand through mental commands. The time-frequency plot in Figure \ref{unfolded_spctr} provides a snapshot of the subject’s efforts, showing the suppression of activity during MI task periods, which corresponds with the robust performance of the real-time BCI. Notably, SMR activity does not remain at a tonic level; instead, it appears as brief spindles recurring intermittently throughout each trial, with gaps between successive spindles. This characteristic introduces inherent latency in MI-BCI systems (especially in simpler implementations), as it necessitates smoothing and accumulating spindle activity for accurate control. A video illustrating the real-time BCI performance can be viewed at external link: https://www.youtube.com/watch?v=oYFgazNzXZM.}

\section{Discussion}
{This study explores the way for next-generation motor-imagery BCIs utilizing portable OPM magnetoencephalography. We conducted experiments in two distinct MSR environments, employing analog and digital suppression systems respectively. Our findings demonstrate that the digital suppression system can effectively transform the intra-MSR environment from one that is unsuitable for experiments to one characterized by optimal background noise levels. This advancement enables the rigorous execution of OPM experiments and facilitates the reproduction of prior results.}

{We observed a diverse array of SMR components in both MI and MR paradigms, each exhibiting a distinct spatio-spectro-temporal profile. The variability in the spectral characteristics of these SMR components can be attributed to several factors, including individual differences among subjects, which encompass variations in cortical architecture, age, anatomy, and other influences. Notably, for subjects who participated in experiments across both rooms, the general patterns of SMR components remained consistent, despite the influence of temporal and sensor configuration factors. As anticipated, MI tasks elicited slightly weaker patterns of MRD, however, these patterns remained sufficiently robust, with SMR profiles closely mirroring those observed in the MR paradigm, differing primarily in MRD magnitude. The time-frequency plots provide a detailed representation of spindle dynamics, revealing such interesting details as the SMR-rebounding effect.}

{We validated the informativeness of the recorded data through offline classification, which in cases with strong MI responses achieved exceptionally high AUC metrics, approaching 1.0 even with a simple linear classification algorithm. Additionally, we investigated the influence of different orientations of the brain’s magnetic field at the head surface on classification accuracy and demonstrated that, for most subjects, the radial axis is likely more informative than the tangential axis. This effect cannot be attributed to differences in OPM sensitivity across axes as they have same sensitivity, suggesting that factors such as the alignment of tangential axes or artifacts from external fields may contribute to this outcome. While an optimal time window and sensor count improve classification performance, moderate accuracy can still be achieved with a single OPM channel. This, along with the successful decomposition of SMR signals from a small recording area, underscores the high spatial resolution capabilities of OPM sensors. Our online BCI paradigm further highlights a key challenge: the latency of algorithmic processing for external device control, which stems from the discrete spindle-like nature of SMR rhythm. Nonlinear approaches may offer a solution, potentially reducing the required window size for feature accumulation in SMR processing.}

{The limitation of this study is the limited number of sensors used, which constrains the ability to capture whole-head activity during motor task, including signals from contralateral hemispheres and additional relevant regions. This restriction also led us to focus on two-class classification rather than exploring multiclass experiments. Future directions include the development of a multichannel, versatile OPM system equipped with artificial active noise control to enhance environmental noise suppression.}

\bibliographystyle{unsrtnat}
\bibliography{references}  






\end{document}